\documentclass[%
 reprint,
 amsmath,amssymb,
 aps,
]{revtex4-2}

\usepackage{acro}
\DeclareAcronym{ANN}{short=ANN, long=Artificial Neural Network}
\DeclareAcronym{DFT}{short=DFT, long=Density Functional Theory}
\DeclareAcronym{GPR}{short=GPR, long=Gaussian Process Regression}
\DeclareAcronym{GAP}{short=GAP, long=Gaussian Approximation Potential}
\DeclareAcronym{GP}{short=GP, long=Gaussian Process}
\DeclareAcronym{SOAP}{short=SOAP, long=Smooth Overlap of Atomic Positions}
\DeclareAcronym{PES}{short=PES, long=Potential Energy Surface}
\DeclareAcronym{QUIP}{short=QUIP, long=Quantum and Interatomic Potentials}
\DeclareAcronym{MLIP}{short=MLIP, long=Machine Learned Interatomic Potential}
\DeclareAcronym{ASE}{short=ASE, long=Atomic Simulation Environment}
\DeclareAcronym{MPI}{short=MPI, long=Message Passing Interface}
\DeclareAcronym{ACE}{short=ACE, long=Atomic Cluster Expansion}
\DeclareAcronym{CEBE}{short=CEBE, long=core-electron binding energy}
\DeclareAcronym{RMSE}{short=RMSE, long=Root Mean Square Error}

\usepackage{graphicx}
\usepackage{dcolumn}
\usepackage{bm}
\usepackage{xfrac}
\usepackage{xspace}
\usepackage{siunitx}
\usepackage{hyperref} % load before cleveref
\usepackage{cleveref}
\usepackage{subcaption}
\usepackage{enumitem}
\usepackage{rotating}
\usepackage{makecell}

\newlist{commalist}{description*}{1}
\setlist[commalist]{
  itemjoin={{,}},
  itemjoin*={{, and}},
  afterlabel=\unskip{{~}}
}

\newlist{arglist}{description}{1}
\setlist[arglist]{
  itemsep=0ex,
  leftmargin=1em,
  labelsep=1em,
  align=left,
  font=\ttfamily\bfseries,
  before={}
}

\newlist{annotatedlist}{description*}{1}
\setlist[annotatedlist]{
  basicstyle=\small\ttfamily,
  columns=fullflexible,
  linewidth=\columnwidth,
  breaklines=true
}

\usepackage[final]{changes}
\usepackage{dsfont}
\usepackage{bm,upgreek}

\newcommand{\kc}{c}

\newcommand{\bb}{\mathbf{b}}
\newcommand{\bA}{\mathbf{A}}

\newcommand{\bc}{\mathbf{\kc}}

\newcommand{\bI}{\mathbf{I}}

\newcommand{\bK}{\mathbf{K}}

\newcommand{\bQ}{\mathbf{Q}}
\newcommand{\br}{\mathbf{r}}
\newcommand{\bR}{\mathbf{R}}
\newcommand{\bx}{\mathbf{x}}
\newcommand{\by}{\mathbf{y}}

\newcommand{\T}{^\top}

\newcommand{\bSigma}{\boldsymbol{\Sigma}}

\newcommand{\etal}{\emph{et al.}}
\newcommand{\abinitio}{\emph{ab initio}}

\newcommand{\code}[1]{\texttt{#1}}

\graphicspath{ {img/} }

\begin{document}

%\preprint{APS/123-QED}

\title{Gaussian Approximation Potentials: theory, software implementation and application examples}

\author{Sascha Klawohn}
\affiliation{Warwick Centre for Predictive Modelling, School of Engineering, University of Warwick, Coventry CV4 7AL, United Kingdom}

\author{G\'abor Cs\'anyi}%
\affiliation{Engineering Laboratory, University of Cambridge, Cambridge CB2 1PZ, UK}

\author{James P. Darby}
\affiliation{Warwick Centre for Predictive Modelling, School of Engineering, University of Warwick, Coventry CV4 7AL, United Kingdom}

\author{James R. Kermode}
\affiliation{Warwick Centre for Predictive Modelling, School of Engineering, University of Warwick, Coventry CV4 7AL, United Kingdom}

\author{Miguel A. Caro}
\affiliation{Department of Chemistry and Materials Science,
Aalto University, 02150, Espoo, Finland}

\author{Albert P. Bart\'ok}
\email[]{apbartok@gmail.com}
\affiliation{Department of Physics, University of Warwick, Coventry CV4 7AL, United Kingdom}
\affiliation{Warwick Centre for Predictive Modelling, School of Engineering, University of Warwick, Coventry CV4 7AL, United Kingdom}
\date{\today}% It is always \today, today,

\begin{abstract}
\Aclp{GAP} are a class of \Aclp{MLIP} routinely used to model materials and molecular systems on the atomic scale.
The software implementation provides the means for both fitting models using \abinitio{} data and using the resulting potentials in atomic simulations.
Details of the \acs{GAP} theory, algorithms and software are presented, together with detailed usage examples to help new and existing users. We review some recent developments to the GAP framework, including MPI parallelisation of the fitting code enabling its use on thousands of CPU cores and compression of descriptors to eliminate the poor scaling with the number of different chemical elements.
\end{abstract}

\maketitle

%\listoftodos[Notes]

\section{Introduction}

\acfp{MLIP} have revolutionised atomic simulations by offering predictive and computationally inexpensive force fields\cite{Behler.2007,ACE_ralf,Bartok.2010,Schutt.2018}.
While their implementations differ, these models approximate the \ac{PES} of atomic systems based on a database of atomic configurations with corresponding high-accuracy properties calculated with \abinitio{} quantum mechanical methods.

\added{Among the numerous related methods and their software implementations\cite{Behler.2007,ACE_ralf, Gubaev.2018,Zhang.2018die,Jinnouchi.20197ve, Batzner.2022,Chen2022}, the \acf{GAP} approach follows a Bayesian approach, which allows the formulation of prior knowledge about the atomic system and interactions as hyperparameters, as well as uncertainty estimation.
A further advantage of using \ac{GPR} is that fitting the model is a convex optimisation problem\cite{Williams.2007e} equivalent to the solution of a linear system, therefore many problems associated with minimising the loss function of neural networks\cite{NIPS2016_f2fc9902} do not occur.}

The \ac{GAP} framework, originally proposed by Bart\'ok \etal{}\cite{Bartok.2010}, uses \ac{GPR} to infer local atomic properties via a set of descriptors that map Cartesian atomic coordinates to invariant representations.
\Ac{GAP} models have been used successfully to model silicon\cite{Bartok.2010,Deringer.2020}, carbon\cite{Deringer.2017}, tungsten\cite{Szlachta.2014}, phosphorus\cite{Deringer.202038}, water\cite{Bartok.2013kgn}, iron\cite{Dragoni.2018}, gold and platinum\cite{kloppenburg_2022b,kloppenburg_2023}, hafnia\cite{Sivaraman.2021} and gallia\cite{Liu.2020v6s}, among others.

\added{Similarly to other \ac{MLIP} frameworks, the \ac{GAP} package can utilise reference atomic databases produced with arbitrary \abinitio{} methods and software packages.
Total energies and derivative quantities (forces and stresses) are used to fit the \ac{PES}, although models for local atomic properties, such as nuclear magnetic resonance (NMR) shieldings or Hirshfeld volumes may also be generated using our software. 
Recently, \ac{GAP} was used to accelerate \abinitio{} molecular dynamics simulations within the CASTEP\cite{CASTEP} package, utilising an adaptive scheme that produces an evolving and improving \ac{GAP} model during the dynamics\cite{10.1063/5.0155621}.}

In this paper, we present the current status of the \ac{GAP} framework, discussing the particular adaptation of the sparse \ac{GPR} that enables a performant implementation suitable to fit energetic properties of atomic systems.
\Ac{GAP}, as implemented within the \ac{QUIP}\cite{QUIPGAPGithub} software package, is introduced, emphasising the flexibility and extensibility of the code.
We also discuss recent developments, such as parallelisation and compression of  descriptors, making connections between the theory and practical usage.
Finally, we present some examples, which are intended to illustrate a selection of features enabled by the \ac{GAP} package.

By documenting implementation details for the available options, this paper is not only intended for practitioners fitting \ac{GAP} models, but also for those developing other \ac{MLIP} frameworks.  

%Solid-state, DFT, configurations, machine learning, Gaussian process (GP), Gaussian Approximation Potential (GAP)\cite{GAP}, sparse/basis set

\section{Theory}

%\todo[inline]{Distinguish between xyz input and descriptor input $\bx$.}

The \ac{GAP} framework utilises sparse \ac{GPR} which is customised to fit \acp{PES} as well as local properties of atomic systems.
A detailed introduction to \ac{GPR} can be found elsewhere\cite{Mackay,Williams.2007e} and background on the \ac{GAP} framework is presented in the review article by Deringer \etal{}\cite{Deringer.2021}.
Here we revise only the formulae necessary for discussing the software implementation.

A central assumption in fitting the \ac{PES} of atomic systems is that the total quantum mechanical energy may be decomposed into local contributions $\varepsilon$ which depend on descriptors $\bx$:
\begin{equation}
    E = \sum_d ^\textrm{descriptors} \sum_{i=1}^{N_d} \varepsilon_d(\bx_i),
\end{equation}
where $N_d$ is the number of descriptors of type $d$.
Descriptors may be the arguments to two-body energy terms, based on the interatomic distance, optionally augmented by the symmetrised atomic coordinations, or three-body terms, based on the bond angle and the symmetrised bond distances, optionally including the coordination of the central atom.
The greatest advantage of the non-linear regression techniques enabled by machine learning methods is the ability to parameterise the highly complex many-body energy terms.
The descriptors forming the arguments to these functions may be $n$-body terms, based on the interatomic distances within a cluster of $n$ atoms, or flexible many-body terms, based on the \ac{SOAP}\cite{Musil.2021} descriptor, the bispectrum\cite{Bartok.2010,Bartok.2013} or the Behler-Parrinello symmetry functions\cite{Behler.2007}.
Finally, \ac{GAP} implements customised descriptors, to represent molecules, dimers and trimers.

In \ac{GAP}, each energy term $\varepsilon_d$ is written as an independent sparse Gaussian Process, in the form
\begin{equation}
  \varepsilon_d(\bx) = \sum_{m=1}^{M_d} c_m k_d(\bx,\bx_m)
    \textrm{,}
    \label{eq:epsilon}
\end{equation}
where $M_d$ is the number of sparse or representative points of descriptor $d$, $k_d$ is the kernel, covariance or similarity function and $c_m$ are the fitting coefficients.

The coefficients in \eqref{eq:epsilon} are fitted using a database of atomic configurations, where corresponding total energies and derivative quantities, such as forces and virial stresses, have been determined using \abinitio{} quantum mechanical calculations.
The target properties, denoted by $\by$, of the fitting procedure are therefore the sum of local energy contributions in the form of total energies, or the sum of the partial derivatives of local energy terms in the form of forces or virial stresses.
The differentiation operator with respect to a Cartesian coordinate $r_{i\alpha}$ is propagated through the kernel function, resulting in partial derivatives
\begin{equation}
    \frac{\partial \varepsilon}{\partial r_{i\alpha}} =
    \sum_{m=1}^{M_d} c_m \nabla_{\bx} k(\bx,\bx_m) \frac{\partial \bx}{\partial r_{i\alpha}}
    \textrm{.}
\end{equation}
The sparse \ac{GPR} adapted to our case\cite{Deringer.2021} becomes
\begin{equation}
\mathbf{c}  = [\mathbf{K}_{MM} + 
    (\hat{L} \mathbf{K}_{NM})\T \bSigma^{-1} \hat{L}\mathbf{K}_{NM}]^{-1} (\hat{L}\mathbf{K}_{NM})\T \bSigma^{-1} \mathbf{y}
    \textrm{.}    
    \label{eq:GP_L}
\end{equation}
The kernel or covariance matrices $\mathbf{K}_{MM}$ and $\mathbf{K}_{NM}$ contain the function values $k(\bx_m,\bx_{m'})$ and $k(\bx_n,\bx_{m})$ respectively, where $m$ and $m'$ denote sparse points and $n$ denote descriptors of the database configurations.
In case of $\mathbf{K}_{NM}$, kernel functions may be the derivative 
 values, $-\nabla_{\bx} k(\bx_n,\bx_m) \frac{\partial \bx_n}{\partial r_{i\alpha}}$, if the corresponding target observation in $\by$ is a force quantity.
The diagonal $\bSigma{}$ matrix contains the regularisation strength parameters \added{($\sigma_{\textrm{energy}}$, $\sigma_{\textrm{force}}$ and $\sigma_{\textrm{virial}}$, which may be specified individually)}, encoding the prior assumption regarding the accuracy of target values.
Finally, the operator $\hat{L}$ is a shorthand for the summation that accumulates the local terms composing each target value in $\by$.

%Given data $\mathcal{D}$ consisting of descriptive input $\bx$ and target properties $\by$ (also $\by'$) our goal is to define and train a model so that for another input $\bx^{*}$ we may calculate the properties $\hat \by^{*}$.

%Kernel matrix $K_{ij} = k (\bx_i, \bx_j)$

%subset of $M \ll N$ sparse points GP\cite{QuinoneroCandela:2005wp,Snelson:2006vi}.

% … leads to
% \begin{equation}
%     \mathbf{c}  = (\mathbf{K}_{MM} + 
%     \mathbf{K}_{MN} \bSigma^{-1} \mathbf{K}_{NM})^{-1} \mathbf{K}_{MN} \bSigma^{-1} \by
%     \label{eq:GP_sparse}
% \end{equation}

Foster \etal{} have shown\cite{Foster:2009wy} that solving equation \eqref{eq:GP_L} directly can lead to numerically unstable results, in which uncertainties in the input lead to disproportionate errors in the output.
Instead, we first define
\begin{equation}
    \mathbf{A} = \begin{bmatrix}
    \bSigma^{-\sfrac{1}{2}} \hat{L}\mathbf{K}_{NM} \\
    \mathbf{U}_{MM}
    \end{bmatrix},
\end{equation}
where the Cholesky decomposition of $\mathbf{K}_{MM}$ results in the upper triangular matrix $\mathbf{U}_{MM}$ such that $\mathbf{K}_{MM} = \mathbf{U}_{MM}^T\mathbf{U}_{MM}$.
While $\mathbf{K}_{MM}$ is positive semidefinite, depending on the database configurations and descriptor types, the sparse points may be highly correlated, leading to an ill-conditioned $\mathbf{K}_{MM}$ matrix, preventing the Cholesky decomposition we use to obtain $\mathbf{U}_{MM}$.
To regularise the sparse covariance matrix $\mathbf{K}_{MM}$, we add a small constant to the diagonal, informally called the \emph{jitter}, which is typically 8-10 orders of magnitude less than the elements of $\mathbf{K}_{MM}$.
\added{
The \emph{jitter} has a similar effect on the resulting sparse model as the noise hyperparameter in a full \ac{GPR}.
As both the aleatoric and epistemic uncertainty is controlled by $\bSigma$, the error in the local energy term introduced by the use of \emph{jitter} is a small broadening, which is expected to be on the order of the square root of the \emph{jitter}.
}

Padding the vector of target properties $\by$ by an $M$-long vector of zeros,
\begin{equation}
    \mathbf{b} = \begin{bmatrix}
        \by \\
        \mathbf{0}
    \end{bmatrix},
\end{equation}
we rewrite \cref{eq:GP_L} as the solution of the least-squares problem
\begin{equation}
    \min_\mathbf{c} (\mathbf{A}\bc - \mathbf{b})^T (\mathbf{A}\bc - \mathbf{b}),
\end{equation}
leading to the solution in the form of
\begin{equation}
    \bc = (\mathbf{A}^T \mathbf{A})^{-1} \mathbf{A}^T \bb
    \textrm{.}
    \label{eq:sparseAsolution}
\end{equation}
A numerically stable solution can be obtained by first carrying out a QR factorisation of $\bA = \bQ \bR$ where $\bQ$ is orthogonal, namely, it is formed by \emph{orthonormal} column vectors, while $\bR$ is an upper triangular matrix.
Substituting the factorised form of $\bA$ into \cref{eq:sparseAsolution} results in
\begin{equation}
    \bc = (\bR^T \bQ^T \bQ \bR)^{-1} \bR^T \bQ^T \bb = \bR^{-1} \bQ^T \bb,
\end{equation}
as $\bQ^T \bQ = \bI$.

\section{Implementation}

We develop and maintain a collection of software tools called QUIP\footnote{Quantum mechanics and Interatomic Potentials} to carry out molecular dynamics simulations.
Part of this suite is an implementation of \ac{GAP}, including the \verb|gap_fit| program, implementing the sparse \ac{GPR}.
%It was initially written in ((object-oriented language, C++???)) in ((year)), then rewritten in Fortran in ((year)).
The majority of \ac{QUIP} is written in modern Fortran, utilising many object-oriented features, although not fully exploiting the \replaced{most recent}{later} Fortran standards due to lack of compiler support at the time of the original development of the code, which started in 2005.
\ac{QUIP} features a Python interface, \code{quippy}\cite{quippy}, allowing access to various functionalities and casting all atomic potentials as \ac{ASE} calculators\cite{Larsen.2017}.
There also exist generic C and LAMMPS-specific C++ interfaces, that allow \ac{GAP} models to be used in external simulation packages.
The source code can be found on GitHub\cite{QUIPGAPGithub}.

\subsection{The GAP submodule}
Similarly to other \acp{MLIP}, the main components of \ac{GAP} are the calculation of descriptors, mapping Cartesian coordinates into invariant representations, and a regression method, in this case \ac{GPR}.
\subsubsection{Representations}
The \verb+descriptors.f95+ file implements the mapping from the Cartesian coordinates of the atoms to invariant representations.
\added{
In the package we provide a number of descriptors, but it is straightforward to implement new ones.
While the user fitting \ac{GAP} models does not have to interact with the source code, in the following we give an overview of what is necessary to implement or adapt a descriptor.
}

A set of standardised interfaces are used for each descriptor, which are overloaded, therefore adding new descriptors is straightforward and does not require any further modifications elsewhere.
The \verb|initialise| interface interprets the parameters of the descriptor, which are provided by the user as key-value pairs, and stores these in a \verb|descriptor| object.
The query function \verb|cutoff| returns the spatial cutoff of a descriptor, whereas \verb|finalise| resets the descriptor object and deallocates all storage.
The \verb|descriptor_sizes| function takes an \verb|Atoms| object and determines how many descriptors and partial derivatives will be calculated based on the cutoff and the connectivity of the particles.
Finally, the \verb|calc| function returns descriptors calculated based on an \verb|Atoms| and a \verb|descriptor| object, and optionally, their partial derivatives with respect to atomic coordinates in a generic container object.
All of this functionality is exposed in \verb|quippy|, ensuring interoperability with \ac{ASE}.

\subsubsection{Regression}
\ac{GPR} is implemented in the file \verb|gp_predict.f95|, with some fitting-specific features in \verb|gap_fit_module.f95| and sparse point selection in \verb|clustering.f95|.
For a pair of descriptors $\bx$ and $\bx'$ of dimensions $D$, whose distance $r$ is defined as
\begin{equation}
    r = \sqrt{\sum_{i=1}^D \frac{(x_i-x_i')^2}{2\theta_i^2}}
    \textrm{,}
\end{equation}
we have implemented the squared exponential kernel
\begin{equation}
    k_{\textrm{SE}}(\bx,\bx') \equiv k_{\textrm{SE}}(r) = \exp (-r^2)
    \textrm{,}
\end{equation}
and the piece-wise polynomial kernel with compact support
\begin{equation}
    k_{\textrm{PP}}(\bx,\bx') \equiv k_{\textrm{PP}}(r) = (1-|\bx-\bx'|)^{j+1} [(j+1)|\bx-\bx'| + 1]
    \textrm{,}
\end{equation}
where $j=\lfloor \frac{D}{2} \rfloor + 2$.
The \ac{SOAP} descriptors should be used with the dot-product or, more generally, polynomial kernel
\begin{equation}
    k_{\textrm{DP}}(\bx,\bx') = (\bx \cdot \bx')^\zeta
    \textrm{.}
\end{equation}

The hyperparameters, such as $\boldsymbol{\theta}$ or $\zeta$, and the coefficients $\bc$ are stored in a Fortran object which is used by the function \verb|gp_predict| to evaluate \eqref{eq:epsilon} as well as the partial derivatives with respect to descriptor components and variances predicted from \ac{GPR}.

During the training procedure, all descriptors $\bx$ and their partial derivatives are calculated and stored.
Pointers are used to denote which descriptors and derivatives contribute to target properties, thereby avoiding the need to store repetitive information.
The kernel matrices $\mathbf{K}_{NM}$ used in the fitting procedure in \eqref{eq:GP_L} are not calculated explicitly, only the accumulated terms in $\hat{L}\mathbf{K}_{NM}$ corresponding to quantum mechanical observable quantities.

\subsection{\ac{GPR} using \texttt{gap\_fit}}

Finding the coefficients used in \ac{GAP} models can be accomplished using the \verb|gap_fit| command line program, where parameters are set as arguments or a configuration file.
The input to the fitting procedure consists of the fitting data, model definitions, and additional options.
The database of atomic configurations, together with the quantum mechanical properties are read in from an extended XYZ\cite{extxyz} file.
The extended XYZ contains information of the lattice, atomic numbers and Cartesian coordinates, and may provide the total energy, forces and virial stress, or any combination of these for each individual configuration.

Each individual atomic configuration may optionally have a type specification, given within the extended XYZ file using the \code{config\_type} keyword.
The configuration type may be used for fine-grain control, such as selecting a specific number of sparse points from those configurations.

The particulars of the model are provided using the \texttt{gap} argument as a form of colon-separated list of descriptor definitions.
These include hyperparameters, such as the spatial cutoff or the desired number of sparse points per descriptor.
Other hyperparameters, such as the regularisation, can be provided either in the command line, or specifically for each individual frame within the extended XYZ file, allowing fine-grained control and the use of inhomogeneous accuracy of target quantities.
Additional options can be used to adjust the processing, tune technical parameters, or enable additional features like more verbose output.
Details about the currently available arguments can be found later in \cref{sec:args}, or running \texttt{gap\_fit --help} for up-to-date information.

\subsubsection{Program structure}

After initialisation and reading of the input, the extended XYZ frames, where each frame consists of an atomic structure, are parsed for the number of target properties and descriptors.
Storage for the descriptor arrays are allocated accordingly, and descriptors are calculated during a second pass over the atomic configurations.
%The target data points are stored in $\by$ and pointers are 

For each set of $N$ descriptors, $M \ll N$ points are chosen as a representative, or sparse, set.
Options include random selection, $k$-means clustering, a uniform grid spanning the range of descriptors and CUR decomposition\cite{Mahoney.2009}.
It is also possible to provide the sparse points via text files.
As the sparse points need to form a linearly independent covariance matrix, duplicates within a given tolerance are removed and only considered once.
This may result in fewer sparse points used than specified by the user, particularly if the atomic environments in the database are highly correlated.

With the specified sparse points, the covariance matrices $\bK_{MN}$ and $\bK_{MM}$ are calculated, and matrix $\bA$ constructed. The coefficients are determined via QR decomposition using (Sca)LAPACK\cite{lapack99,scalapack97} routines.

The memory requirement for the \verb|gap_fit| program  depends on the atomic structures, the number of target properties and sparse points, and descriptor definitions.
In particular, the two main data components with significant memory requirements are the descriptors and their partial derivatives
and the kernel matrix $\hat{L}\mathbf{K}_{NM}$.
The memory associated with storing descriptor derivatives scales linearly with the number of atomic environments and the dimensionality of the descriptor, as well as the number of neighbours within the spatial cutoff.
Efforts directed at developing compact descriptors, using compression techniques, therefore significantly reduce the memory requirements of the fitting procedure, as well as the computational complexity of evaluating the descriptors.
The size of the kernel matrix scales linearly with the number of target values and the total number of sparse points.

\begin{figure}
    \includegraphics[width=0.8\columnwidth]{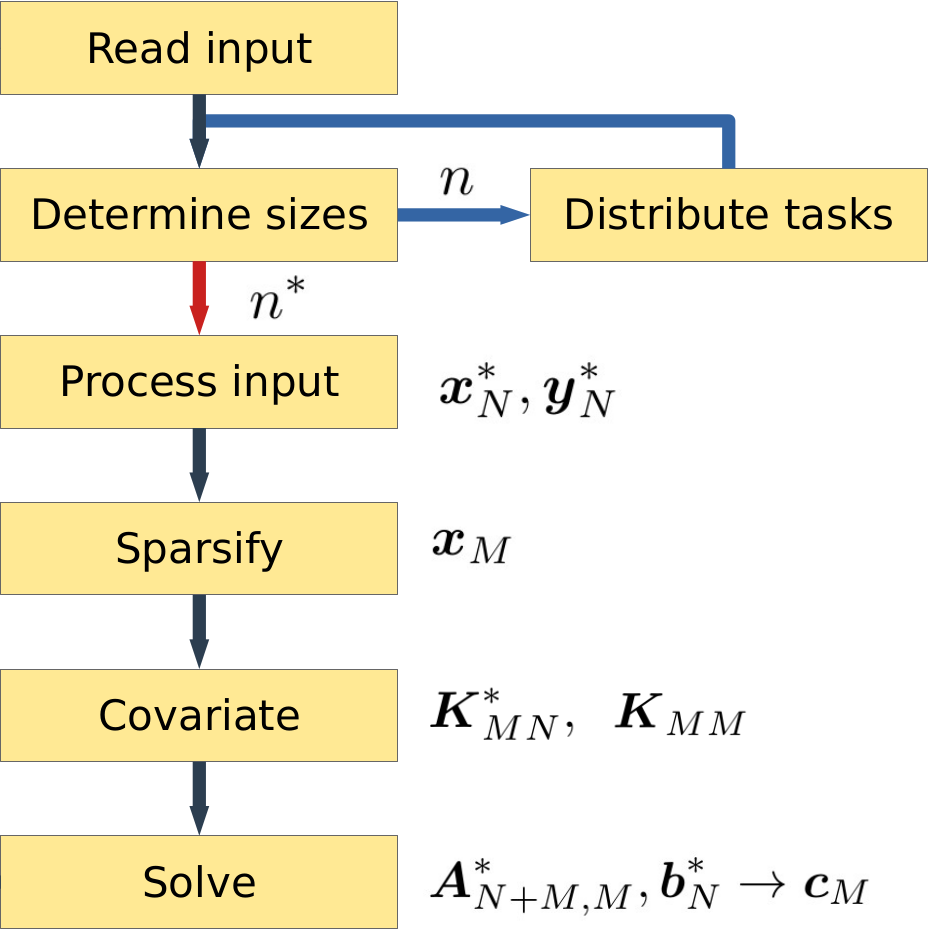}
    \centering
    \caption{MPI \texttt{gap\_fit} workflow. If run in serial, the tasks are not distributed and $n$ is used instead of $n^*$.}
    \label{fig:mpi}
\end{figure}

Recently, we have implemented domain decomposition in \verb|gap_fit| that aims to utilise massively parallel computer architectures\cite{Klawohn.2023}.
The implementation relies on \ac{MPI} and is illustrated on Fig.~\ref{fig:mpi}.

\begin{figure}
    \includegraphics[width=\columnwidth]{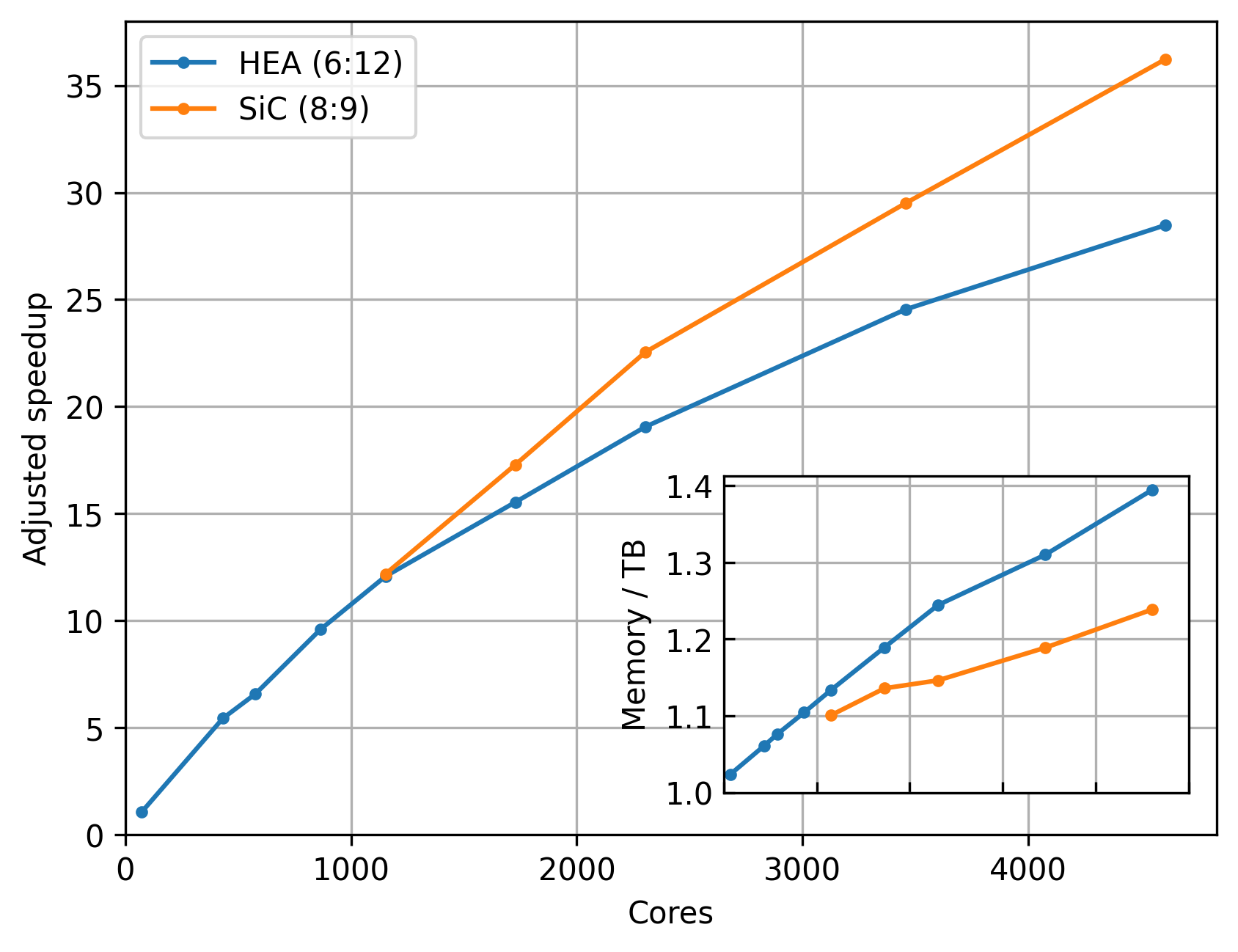}
    \centering
    \caption{Speedup (main panel) and memory (inset) requirements for fitting a high-entropy alloy (HEA) and silicon carbide (SiC) training set using a varying number of nodes.}
    \label{fig:hea_sic}
\end{figure}

After determining the number of atoms and, consequently, the number of target data values in each configuration of the database, configurations are assigned to individual \ac{MPI} tasks.
Descriptors are computed locally, and the covariance matrix $\bK_{MN}$ is constructed in a distributed fashion.
This approach allows the memory requirements of the program to be distributed over many computational nodes, therefore larger databases can be easily fitted without the need of specialised, high-memory servers.
The linear algebra step makes use of the ScaLAPACK library, carrying out the QR decomposition and subsequent back-substitution steps in parallel, thereby reducing the computational time.
We demonstrate the benefits of the parallel fitting approach by studying two fitting problems that would require significant amounts of wall-time and memory using a single node.
One of the training databases consists of the high-entropy alloy configurations by Byggmastar \etal{}\cite{Byggmastar.2021}, and the other is a data set of silicon carbide configurations from Ref.~\cite{Klawohn.2023}.
The dependence of computational speedup and total memory use is presented on Fig.~\ref{fig:hea_sic}, showing that we have achieved excellent parallel performance, and we can utilise the aggregate memory of multiple computational nodes, thereby largely eliminating any limitations.
We note that the dependence of the memory usage on the number of cores is due to overheads associated with repeated data allocations that are private to a process.
Given the typical amount of memory, on the order of hundred GBs, available on commodity computing nodes, this is not likely to be a significant barrier to large-scale parallel fitting calculations.
For more information, we refer the user to our prior work\cite{Klawohn.2023}, where we explored hybrid OpenMP-MPI parallel strategies that optimise overall memory use and runtime.

% A single node usually does not have more than a few terabytes of RAM, often much less than that. This restricted the broad application of QUIP/GAP to smaller models, training sets, and/or special high memory nodes. To alleviate this restriction, Gap Fit can be run with MPI in a SPMD\footnote{single program, multiple data} paradigm using the ScaLAPACK library for linear algebra computation on distributed data.

\subsubsection{Sparse point selection}

The program \code{gap\_fit} implements several methods for the sparse point selection, controlled by the \code{sparse\_method} command line argument.
Within the definition of each descriptor, the \code{n\_sparse} argument controls the number of sparse points.
Alternatively, \code{config\_type\_n\_sparse} allows the user to specify a given number of sparse points from labelled configurations, to ensure adequate representation.
Given \code{none}, no sparsification is applied, apart from the removal of duplicate descriptors, and all points are selected.
The points may be chosen directly with either the \code{file} or \code{index\_file} option, in conjunction with the \code{sparse\_file} argument to specify the filename containing the descriptors or the indices of descriptors.
The indices are 1-based and refer to descriptors as calculated in the database file.

Recommended choices are \code{uniform} for a \code{distance\_2b} descriptor and \code{cur\_points} for a \code{soap} descriptor.
For completeness, we list all currently implemented options in the Appendix.

\subsection{Descriptors}

The choice of invariant representation of atomic environments has a profound effect on the quality of the resulting interatomic potential.
QUIP, being a test bed for methodological developments of \acp{MLIP}, implements numerous descriptors, of which some frequently used ones are presented in this section.

In the \code{gap\_fit} program, descriptors are specified in the \code{gap} command line argument, using the syntax
\code{gap=\{descriptor1 key=value ... : descriptor2 key=value ...\}}.
The descriptor definitions are, by default, treated as templates by \code{gap\_fit}, and after parsing the database configurations, each descriptor is expanded with element (chemical species) information.
If \code{add\_species=F} is added to the descriptor definition, the descriptor is not modified in this step.

Most descriptors implement the \code{cutoff} keyword, specifying the spatial cutoff within atomic connectivities are considered.
The \code{cutoff\_transition\_width} keyword provides a smooth transition ensuring a numerically well behaved characteristic when the descriptor is used to build an interatomic potential.

\subsubsection{Descriptors based on interatomic distances}
Based on the idea of the cluster expansion of the total energy
\begin{equation}
    E = \sum_i E^{(1)}_i + \sum_{i<j} E^{(2)}_{ij} + \sum_{i<j<k} E^{(3)}_{ijk} + \ldots + E^{(N)}
    \textrm{,}
\end{equation}
the $n$-body terms may be fitted using \ac{GPR} or other regression methods.
The cluster of $n$ atoms, representing the input variable of each term, is well defined by a monotonic function, which could be just the identity, of interatomic distances $\br = [r_{12},r_{13},\ldots]$.

In case of the two-body descriptor, \ac{GAP} implements a polynomial transformation that generates a descriptor from the pair distance $r$ in the form of $[r^{p_1},r^{p_2},\ldots]$, where $\{p_i\}_{i=1}^n$ are a set of exponents.
When using this descriptor in conjunction with a dot-product kernel, the generalised form of a pair potential 
\begin{equation}
    V(r) = \sum_{i=1}^n c_i r^{p_i}
\end{equation}
may be recovered.

However, for three- or higher body energy terms the descriptor formed as a list of interatomic distances is not invariant with respect to the permutation of indices of the same elements within the cluster, therefore cannot be directly used for regression.
Permutational invariance is achieved by symmetrising, then normalising, the kernel:
\begin{align}
    k'(\br,\br') &= \sum_{\hat{P}} k( \br, \hat{P}\br'),\\
    k''(\br,\br') &= \frac{k'(\br,\br')}{\sqrt{k'(\br,\br) k'(\br',\br')}},
\end{align}
where $\hat{P}$ represents the permutation of the order of atoms, and $k''$ is used in the \ac{GPR}.

In \ac{GAP}, we implemented the \code{distance\_nb} descriptor, where the body order is defined using the \code{order} keyword.
The \code{compact\_cluster} keyword specifies the topology of the cluster.
If \code{compact\_cluster=T}, each atom in an atomic configuration is considered as a central atom, and clusters are formed with its $n-1$ neighbours that are within the spatial cutoff.
With the \code{compact\_cluster=F}, all possible graphs where the graph edge has a distance less than the cutoff are formed, allowing, for example, linear chains.

The special cases corresponding to two- and three-body terms are implemented as \code{distance\_2b} and \code{angle\_3b}, respectively.
In case of \code{angle\_3b}, the trimer of atoms formed by the central atom $i$ and its two neighbours $j,k$ is represented by the invariant descriptors $[r_{ij}+r_{ik},(r_{ij}-r_{ik})^2,r_{jk}]$.

As the interatomic distances are used in the kernel directly in these descriptor classes, the implementation of a smoothly varying spatial cutoff must be implemented in the kernel function.
We modify the kernel by multiplying it by a cutoff function which smoothly interpolates between zero, where any of the interatomic distances are greater than the spatial cutoff, and one.
The elementary cutoff function
\begin{equation}
    f_\textrm{cut}(r)=
    \begin{cases}
        1 & \textrm{ if } r < r_\textrm{cut} - d\\ 
        0 & \textrm{ if } r \geq r_\textrm{cut}\\
        \frac{1}{2}\left[ \cos\left(\pi\frac{r-r_\textrm{cut}+d}{d} \right) + 1\right] & \textrm{ otherwise}
    \end{cases}
    \textrm{,}
\end{equation}
where $r_\textrm{cut}$ is the spatial cutoff and $d$ is a transition width, is evaluated for each pairwise distance.
The final cutoff function is obtained as a product of elementary cutoff functions, ensuring that each energy term vanishes smoothly.

\ac{GAP} also allows further adjustment of the tail behaviour of the two-body descriptor, in order to approximate the polynomial decay of some interaction types.
This is achieved by further multiplying the cutoff function by $\left(\frac{\mathrm{erf}(\alpha r)}{r}\right)^q$, where $\alpha$ is a range parameter and $q$ is an exponent.

\subsubsection{SOAP descriptors}
The \ac{SOAP} descriptors were proposed a decade ago\cite{Bartok.2013} as invariant descriptors of atomic environments, and have been used successfully to develop interatomic potentials\cite{Bartok.2018,Deringer.2017,Byggmastar.2021,Deringer.2020}, clustering\cite{De.2016} as well as other machine learning tasks, such as the ShiftML model used to predict Nuclear Magnetic Resonance chemical shifts\cite{Paruzzo.2018}.
For details on the theory, the review by Musil \etal{}\cite{Musil.2021} may be consulted. Here we only repeat what is necessary to explain the implementation options.

To construct the \ac{SOAP} descriptor, the atomic environment is first written as a neighbourhood density function, where atoms of element $\alpha$ are represented by Gaussians:
\begin{equation}
\rho^\alpha(\mathbf{r}) = \sum_i \delta_{\alpha z_i } \exp{\left[\frac{-|\mathbf{r}-\mathbf{r}_i|^2}{2\sigma^2}\right]} f_{\text{cut}}(|\mathbf{r}_i|)
\textrm{,}
\label{eqn:atomic_densities}
\end{equation}
\added{where the sum over $i$ includes all neighbouring atoms with position vector $\mathbf{r}_i$, $z_i$ is the corresponding atomic number and $\sigma$ is a length scale hyperparameter.} The \added{element} density \deleted{channel} $\rho^\alpha$ is expanded in a basis set consisting of the products of orthonormal radial basis functions $g_n$ and the spherical harmonics $Y_{lm}$,
\begin{equation}
\rho^\alpha(\mathbf{r}) = \sum_{nlm} c^\alpha_{nlm} Y_{lm}(\mathbf{\hat{r}}) g_n(r),
\end{equation}
resulting in the coefficients $c^\alpha_{nlm}$.
\added{
In the following, we often refer to grouping of certain indices in the basis expansions as \emph{channels}, a term borrowed from signal processing.
Therefore, the element indices $\alpha$ are the \emph{element channels} and the radial basis indices $n$ are the \emph{radial channels}.
The spherical harmonics indices $l$, together with the corresponding $m$ indices, form the \emph{angular channels}.
}

An invariant kernel or similarity function of two atomic environments is obtained by calculating the overlap of the densities, which has to be integrated over all rotations:
\begin{equation}
    k(\rho, \rho') = \int \text{d}\hat{R} \left | \sum_\alpha \int \text{d}\mathbf{r} \rho^\alpha(\mathbf{r}) \rho'^\alpha(\hat{R}\mathbf{r})\right |^\nu
    \label{eqn:full_kernel}
    \textrm{.}
\end{equation}

For the choice of $\nu=2$, the \ac{SOAP} kernel can be analytically evaluated in the form of a dot-product kernel
\begin{align}
    k(\rho, \rho') &= \sum_{\alpha\beta}\sum_{nn'l} p^{\alpha\beta}_{nn'l} p'^{\alpha\beta}_{nn'l} = \mathbf{p} \cdot \mathbf{p'}, \\
    p^{\alpha\beta}_{nn'l} &= \sum_m c^{\alpha*}_{nlm} c^{\beta}_{n'lm} =  \mathbf{c}_{nl}^{\alpha*} \cdot \mathbf{c}^\beta_{n'l}
    \label{eqn:MS_powerspec}
    \textrm{,}
\end{align}
due to the properties of the Wigner D-matrices representing the rotational transformation of the coefficients.
\added{To ensure that $\bar{k}(\rho,\rho)=1$ for any $\rho$, we normalise the kernel as}
\begin{equation}
    \bar{k}(\rho,\rho') = \frac{k(\rho,\rho')}{\sqrt{k(\rho,\rho)}\sqrt{k(\rho',\rho')}}
    \rm{.}
\end{equation}

The \replaced{dimension}{length} of $\mathbf{p}$ scales as $\mathcal{O}(n_{\max}^2 l_{\max} S^2)$ where $n_{\max}$, $l_{\max}$ and $S$ are the number of radial basis functions, spherical harmonics and elements, respectively.
Apart from the original implementation of \ac{SOAP}, a more efficient variant introduced by Caro\cite{caro_2019} is available as \code{soap\_turbo}.
\added{This descriptor is further
described in Sec.~\ref{sec:soapturbo} and a comparison with regular SOAP is
provided in Sec.~\ref{sec:localproperty}.}

Increasing $n_{\max}$ and $l_{\max}$ improves the resolution of the basis set expansion, and are therefore convergence parameters of the \ac{SOAP} kernel.
Optimal values are strongly dependent of the typical number of neighbours and the Gaussian broadening parameter $\sigma$.
In many applications, the user has the choice to adjust $n_{\max}$ and $l_{\max}$ to achieve the desired balance of accuracy and computational speed.
However, the length of $\mathbf{p}$ has a quadratic dependence on the number of elements, thereby the computational cost of both the components of $\mathbf{p}$ and the $k(\rho, \rho')$ as a dot product are impacted.
Various strategies to reduce this scaling have been proposed, which are discussed below.

\subsubsection{SOAP Compression}\label{sec:compression}

% Increasing $N$ and $S$ increases the cost of computing $\mathbf{p}$ itself \textbf{and} subsequently computing $k(\rho, \rho')$ as a dot product.
\added{
The $\mathcal{O}(l_{\max} n_{\max}^2 S^2)$ scaling of the number of descriptor components in \ac{SOAP} is often limiting as it makes studying chemically diverse systems, such as multi-component alloys
or proteins, very computationally demanding. A widely used approach \cite{willatt2018feature, goscinski2021optimal, gubaev2019accelerating, ACE_ralf, bochkarev2022efficient} to reduce this scaling is
to embed the elements (and optionally radial) channels into a fixed $K$-dimensional space as $\mathfrak{c}^k_{nlm} = \sum_{\alpha} w^k_\alpha c^\alpha_{nlm}$ (or $\mathfrak{c}^k_{lm} = \sum_{n\alpha}
w^k_{n\alpha} c^\alpha_{nlm}$) and then form a compressed descriptor as} 
\begin{equation}
 p^{kk'}_{nn'l} = \sum_m \mathfrak{c}^k_{nlm} \mathfrak{c}^{k'}_{n'lm}.
\end{equation}
\added{
(or $ p^{kk'}_{l} = \sum_m \mathfrak{c}^k_{lm} \mathfrak{c}^{k'}_{lm}$) which reduces the scaling to $\mathcal{O}(l_{\max} n_{\max}^2 K^2 )$ (or $\mathcal{O}(l_{\max}K^2)$). To achieve good performance for $K < S$\cite{lopanitsyna2023modeling} the embedding weights are typically optimised during fitting and, following Willatt \etal{}, they are interpretable as encoding similarity between different elements via the alchemical kernel \cite{willatt2018feature} $\kappa_{\alpha\beta} = \sum_k w^k_\alpha w^k_\beta = \mathbf{w}^\alpha \cdot \mathbf{w}^\beta$. 
}

\added{
 This idea was extended by Darby \etal{}\cite{darby2022tensor}, where it was shown that it is sufficient to couple the embedding channels to themselves only, rather than taking a full tensor product across the embedded index $k$, thus making the scaling linear in $K$, rather than quadratic. Two flavours of these tensor-reduced descriptors were proposed. The first is motivated by considering fitting a linear model as 
}%
\begin{align}
        \varphi &= \sum_{\alpha \beta n n'l} a^l_{(\alpha,n), (\beta,n')} p^{\alpha \beta}_{n n' l} ,
        \label{eq:linear_soap}
\end{align}
\added{where the $a$ are the model coefficients and the element and radial indices have been grouped together. For each value of $l$ the matrix of coefficients $a^l$ can be approximated using
symmetric eigen-decomposition as}
\begin{equation}
    a^l_{(\alpha n), (\beta n')} = \sum_{k=1}^{K} \lambda^l_k w^k_{(\alpha, n)} w^k_{(\beta, n')} 
    \label{eq:eigen-decomposition}
\end{equation}

\added{
This decomposition is exact for $K=n_{\max}S$ with $w$ the eigenvectors of $a^l$ and is systematically improvable with random $w$. Substituting this approximation into Eq. (\ref{eq:linear_soap})
results in} 
\begin{align}
    \varphi &= \sum_{\alpha \beta n n' l} \sum_{k} \lambda^l_k w^k_{(\alpha, n)} w^k_{(\beta, n')} p^{\alpha \beta}_{n n' l} \\
   &= \sum_{k, l}\lambda^l_k  \left(\sum_{\alpha n} w^k_{(\alpha, n)} \bm{c}^\alpha_{nl} \right) \cdot \left(\sum_{\beta n'} w^k_{(\beta, n')} \bm{c}^\beta_{n'l} \right) \\
   & = \sum_{kl} \lambda^l_k \bm{c}^k_l \cdot \bm{c}^k_l= \sum_{kl} \lambda^l_k \tilde{p}^k_l ,
\end{align}
\added{where $\tilde{p}^k_l$ are the new features. As the approximation in Eq. (\ref{eq:eigen-decomposition}) is systematic, any function that can be fit as a linear function of $p^{\alpha \beta}_{nn'l}$ can also be fit using a linear function of $p^k_l$.
}

\added{
An alternative and complementary view motivated by using random mixing weights $w^k_{(\alpha n)}$ is to ``sketch'' the power spectrum as }
\begin{equation}
    \hat{p}^k_l = \left(\sum_{\alpha n} w^k_{\alpha n} \bm{c}^\alpha_{nl} \right) \cdot \left(\sum_{\beta n'} u^k_{\beta n'} \bm{c}^\beta_{n'l} \right)
\end{equation}
so that
\begin{align}
    \mathbb{E}\left[\mathbf{\hat{p}}\cdot \mathbf{\hat{p}}'\right] &= \sum_{kl} \sum_{\substack{\alpha \beta n n' \\ \delta \gamma q q'}} \underset{\sigma^4 \delta_{(\alpha, n), (\delta, q)} \delta_{(\beta, n'), (\gamma, q')} }{\underbrace{\mathbb{E}\left[w^k_{\alpha n} u^k_{\beta n'} w^k_{\delta q} u^k_{\gamma q'}\right]}} \bm{c}^\alpha_{nl} \bm{c}^\beta_{n'l} \bm{c'}^\delta_{ql} \bm{c'}^\gamma_{q'l} \\
    &= \sigma^4 \sum_{k} \sum_{\alpha \beta n n' l} \left(\bm{c}^\alpha_{nl} \cdot \bm{c}^\beta_{n'l} \right) \left(\bm{c'}^\alpha_{nl} \cdot \bm{c'}^\beta_{n'l} \right) \\
    &= K \sigma^4 \mathbf{p} \cdot \mathbf{p'},
\end{align}
\added{
where we have used the fact that $\mathbb{E}[w^k_i w^k_j] = \sigma^2 \delta_{ij}$ if the $w^k_i$ are symmetric random variables with zero mean and that $u$ and $w$ are independent. This is a form of tensor-sketching~\cite{woodruff2014sketching} and is also systematic with the expected error in approximating the kernel decreasing as $K^{-\frac{1}{2}}$.
}

\added{
The different flavours of element-embedding and tensor-reduction listed above are all accessible via specifying various combinations of \texttt{R\_mix}, \texttt{Z\_mix}, \texttt{sym\_mix} and \texttt{K}, which specify how the initial channels should be mixed, and the \texttt{coupling} keyword which specifies how the resulting channels should be coupled together. Note that optimisation of the embedding weights $w$ is not available in \texttt{gap\_fit} with normallly distributed random weights used instead. Please see the keyword glossary for more details.
}

\begin{figure}[h!]
\begin{center}
    \includegraphics[width=0.95\columnwidth]{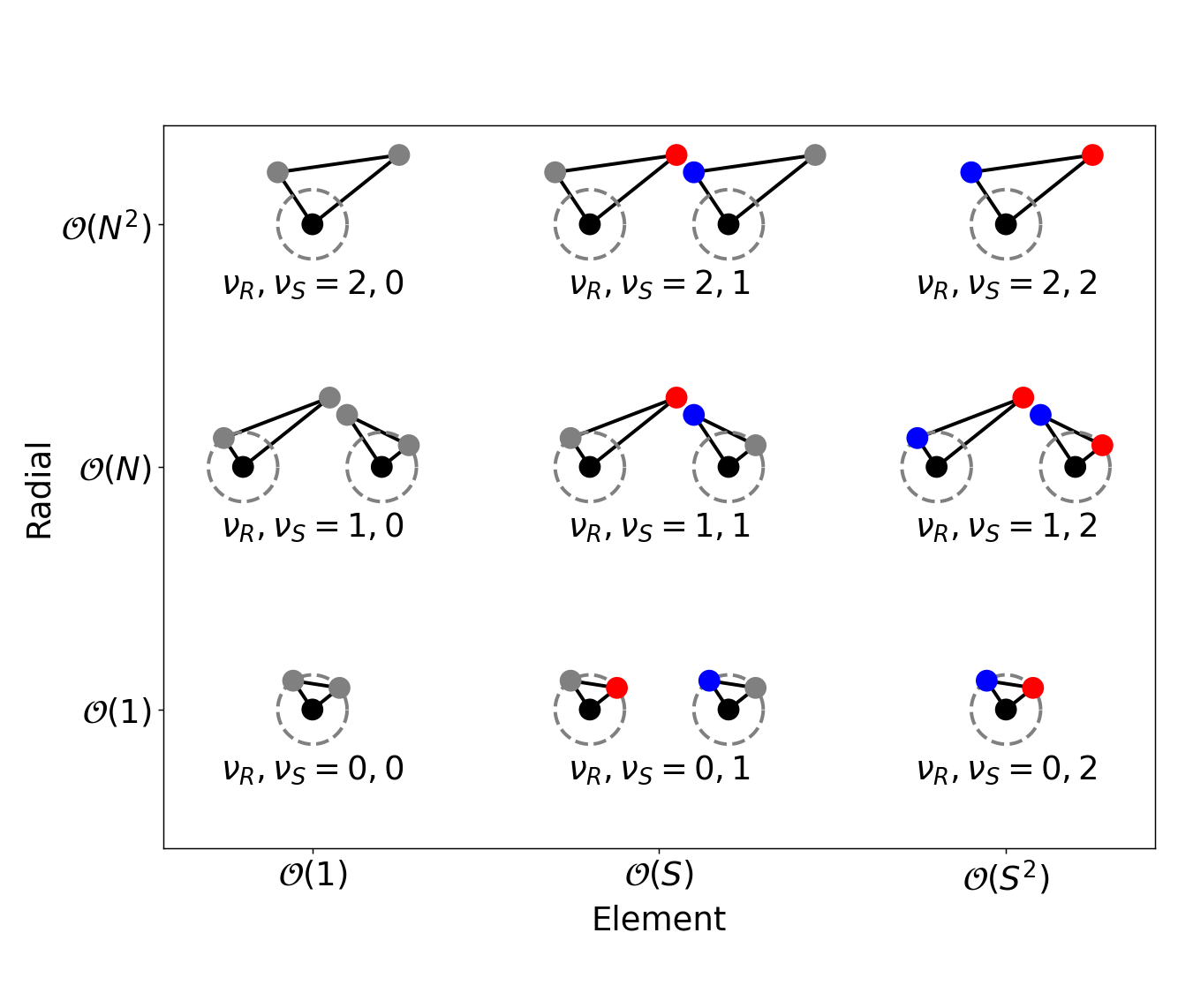} 
\end{center}
    \caption{Different SOAP compression strategies. Neighbour atoms around the central atom (black) may be represented as element-agnostic (grey) or element-specific (red or blue). To eliminate the radial dependence, neighbours may be projected on the unit sphere (dashed circle) around the central atom. Reprinted from Darby
    \textit{et al}\cite{darby2022compressing} with permission under the
    CC BY 4.0 license.}
    \label{fig:nu}
\end{figure}

\added{
An alternative  compression strategy proposed by Darby \etal{}\cite{darby2022compressing} simply involves summing over one (or more) of the $\alpha$, $\beta$, $n$ or $n'$ indices in Eq.
\eqref{eqn:MS_powerspec}. It is efficient to perform this summation at the level of the density expansion coefficients ${c}_{nl}^{\alpha}$ where it can also be most easily understood; summing over a
radial index $n$ is equivalent to projecting the 3D density onto the surface of the unit sphere whilst summing over the element index $\alpha$ corresponds to forming the total, element-agnostic
density. The power-spectrum is a generalised 3-body descriptor where each term in the following sum}
\begin{equation}
    p^{\alpha\beta}_{nn'l} = \sum_{ij} \mathbf{c}_{i, nl}^{\alpha} \cdot \mathbf{c}_{j, n'l}^{\beta}
\end{equation}
\added{
corresponds to a triangle formed by the central atom and the neighbour atoms $i$ and $j$. As such, the various possible effects of this compression scheme on an individual 3-body (correlation order 2)
term in this summation can be visualised as in Fig.~\ref{fig:nu}. The different options are labelled according to the element-sensitive $\nu_S$ and radially-sensitive $\nu_R$ correlation orders where each summation over an element (or radial) index lowers the respective correlation order by one, e.g., $\nu_S=1$, $\nu_R=1$ specifies $p^{\alpha}_{nl} = \sum_{\beta n'} p^{\alpha \beta}_{nn'l}$.
}

\added{
Finally, it is also possible to achieve compression through the experimental \texttt{Z\_map} keyword, which allows the user to group different elements together; equivalent to element embedding with $w^k_\alpha =1$ if element $\alpha$ is in group $k$ or 0 if it is not. As two densities are coupled, two distinct sets of groups may be specified if desired. Please see the keyword glossary for more details.
}

\subsubsection{\code{soap\_turbo} descriptors}\label{sec:soapturbo}

%\textcolor{red}{I'm giving the description of soap\_turbo here, you can reshuffle this info as you see fit}

The \texttt{soap\_turbo} descriptor is a variant of SOAP optimised for computational efficiency.
\added{A detailed account of this descriptor has been given in Ref.~\cite{caro_2019}. Here, we briefly
describe its main features while giving a more in-depth description of the features that have been
introduced since the publication of the original paper, namely multispecies support and compression. A comparison between the \texttt{soap} and \texttt{soap\_turbo} implementations of SOAP is
given in Sec.~\ref{sec:localproperty}.}

The representation of the atomic density field in the local neighborhood, that is, within a cutoff sphere of radius $r_\text{cut}$
of atom $i$, is carried out in an explicitly separable form of radial and angular channels.
Therefore the expansion coefficients can also be split into components that depend exclusively on the
radial index $n$ or angular indices $l,m$:
\begin{align}
\rho_i (\textbf{r}) = \sum\limits_{j \in S_i(r_\text{cut})} \sum\limits_{nlm}
c^{i,j}_{nlm} g_n(r) Y_{lm} (\theta, \phi), \\
c^{i,j}_{nlm} = b^{i,j}_{n} a^{i,j}_{lm},
\end{align}
where $b^{i,j}_{n}$ are the radial expansion coefficients, $a^{i,j}_{lm}$ are the angular
expansion coefficients and $j$ runs over all neighbours of $i$ within the cutoff sphere.
A number of smoothing and scaling functions are introduced, to
make the width of the atom-centered smooth functions depend on the distance from the center of
the SOAP sphere. \added{The main implementation differences between \texttt{soap} and \texttt{soap\_turbo}
are the use of smoother polynomial radial basis functions and several numerical tricks that allow to
express the radial and angular expansion coefficients as recursive series. There are also differences
in how multispecies support and compression are handled, described below.}

Support in \texttt{soap\_turbo} for multiple chemical elements is provided by augmenting the radial basis set
via a direct sum:
\begin{align}
\{g(r)\} = \bigoplus\limits_{s = 1}^{N_\text{s}} \{g(r)\}_s,
\end{align}
where $s$ runs over the number of elements.
The only advantage of this approach compared to the
regular SOAP multielement support is that each element can be represented with a different
radial basis set, including a different number of radial basis functions.
One instance where this feature may be useful is when one of the elements can be represented with fewer radial basis
functions than another, reducing the dimensionality of the descriptor and thus its computational cost.
In principle, this approach also allows for using different cutoff radii for different elements
within the same descriptor although, in practice, the GAP interface currently restricts the cutoff to
be the same for all elements.
The angular basis, on the other hand, is the same for all elements.

Compression is also supported by \texttt{soap\_turbo} through three different approaches.
The first compression scheme is a heuristic recipe, that we refer to as ``trivial'',
which retains only the SOAP elements that run over $n=1$ for
single element and the first radial component of the element-specific basis for multiple elements:
\begin{align}
\{\tilde{p}\} \equiv \left( \bigoplus\limits_{nn'l} \delta_{1,n} p_{nn'l} \right)
\oplus \left( \bigoplus\limits_{nn'l} \delta_{N_\text{r}^1+1,n} p_{nn'l} \right)
\nonumber \\
\oplus \cdots \oplus \left( \bigoplus\limits_{nn'l} \delta_{N_\text{r}^{N_\text{s}-1}+1,n} p_{nn'l} \right),
\end{align}
where the tilde indicates the compressed descriptor and $N_\text{r}^1$ is the number of
radial basis functions for the first element, and the direct sum continues until the
last element in the descriptor has been considered.
The Kronecker deltas indicate that only components with $n=1$, $n=N_\text{r}^1+1$, etc., are retained.
Trivial compression affords of the order of a factor of 5 in dimensionality
reduction without significant loss in accuracy for most production GAP models that we have fitted
so far.

The second compression scheme in \texttt{soap\_turbo} provides a quasiequivalence with the
radial- and element-sensitive correlation orders offered in regular SOAP compression introduced in Section~\ref{sec:compression}. Numerical comparisons of these compression recipes are given in the local property example in Sec.~\ref{sec:examples}.

The third compression scheme has no predefined recipe.
Instead, the user can provide an arbitrary linear transformation (via an input text file)
that projects the SOAP vector from its original $N$-dimensional space to a reduced $M$-dimensional space, where $M<N$:
\begin{align}
\tilde{\textbf{p}} = P \textbf{p}, \qquad P \in \mathcal{M}_{M \times N}.
\end{align}
In all cases, the descriptor is renormalized after compression.

Finally, we remark that due to the overlap properties of the polynomial
radial basis sets and related instabilities in the numerical approach
employed to construct the orthonormal radial basis used to construct
\texttt{soap\_turbo} descriptors, there is a practical limitation of
$n_\text{max} \approx 10$. For most practical purposes (e.g., in
constructing accurate GAP force fields), there is no
need to increase the size of the radial basis set beyond $\approx 8$
basis functions.

\section{Practical examples}\label{sec:examples}
\subsection{Si interatomic potential}

\begin{figure*}
\includegraphics[width=\textwidth]{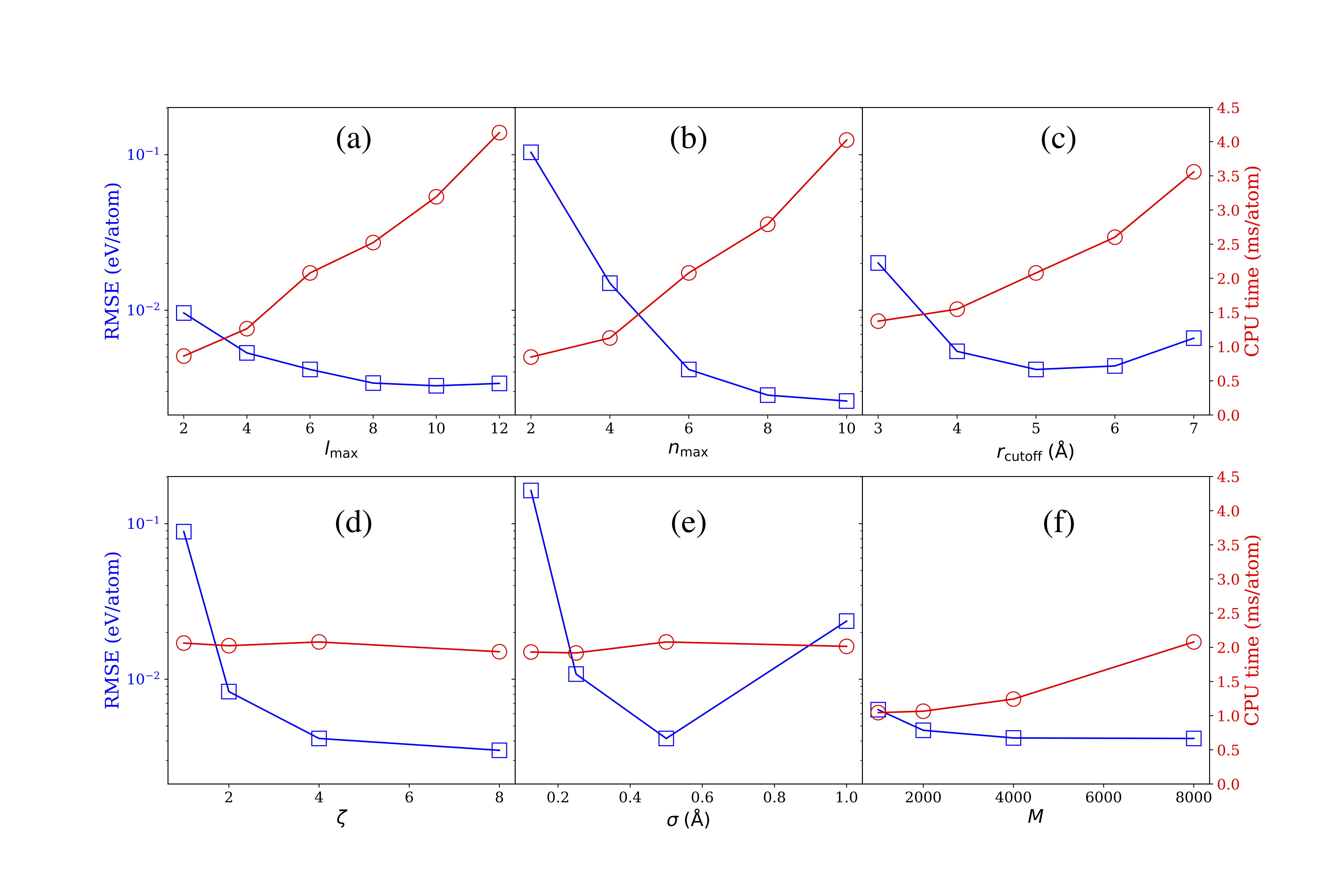} 
\caption{
\added{Performance (blue squares) and computational cost (red circles) of different GAP models.
The performance is quantified by the \ac{RMSE} of the predicted energy, evaluated on a test set.
Comparisons with respect to changes in (a) the radial resolution $n_{\max}$ of SOAP; (b) the angular resolution $l_{\max}$ of SOAP; (c) the spatial cutoff $r_{\textrm{cutoff}}$ of SOAP; (d) the power of the polynomial kernel $\zeta$; (e) the width $\sigma$  of Gaussians representing the atoms in SOAP; (f) the number of sparse points $M$.}
}
\label{fig:Si}
\end{figure*}

\begin{figure}[tb]
\includegraphics[width=\columnwidth]{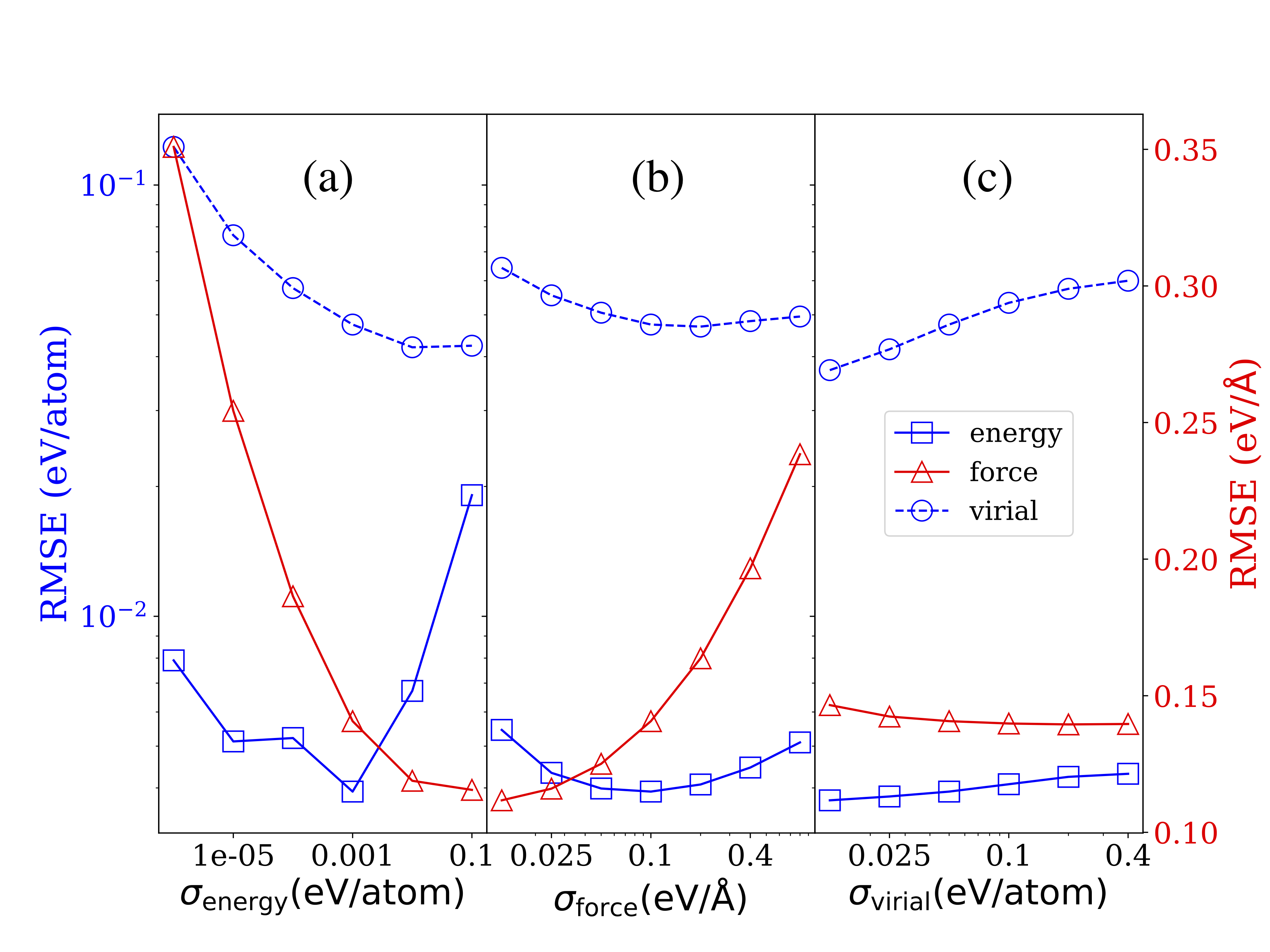} 
\caption{Performance of \ac{GAP} models as the function of regularisation hyperparameters. The RMSE of the predicted energies (blue squares), forces (red triangles) and virial stresses (blue circles) are shown as the energy (panel a), force (panel b) and virial (panel c) regularisation hyperparameters are varied independently.}
\label{fig:sigma_trend}
\end{figure}

\added{
Among the first successful applications of \ac{GAP} was a general-purpose interatomic potential for silicon\cite{Bartok.2018}.
We have used the database of atomic configurations to train a series of \ac{GAP} models to demonstrate the effect of the most crucial descriptor and kernel hyperparameter choices on the performance and computational cost of the resulting potential.
The extended XYZ file, containing the database and shared in the supplementary material of Ref.~\cite{Bartok.2018}, was randomly split into a training and a test set, containing 80\% and 20\% of the original configurations, respectively.
We list the parameters used in the \code{gap\_fit} command line in Table~\ref{tab:configsi} with a detailed explanation for each.
}

\begin{table*}[!t]
\centering
\caption{Command line parameters of \code{gap\_fit} used to fit a \ac{GAP} model for silicon.} 
\label{tab:configsi}
\begin{tabular}{lll}
\hline
\textbf{Key} & \textbf{Value} & \textbf{Comments} \\
\hline
\code{atoms\_filename} & \code{train.xyz} & Extended XYZ file of training configurations \\
\code{gap\_file} & \code{gp\_c5.0\_n6\_l6\_s0.5\_z4\_p8000.xml} & Filename of output XML of \ac{GAP} model\\
\code{energy\_parameter\_name} & \code{dft\_energy} & \makecell[tl]{Target energy key in the extended XYZ file. \\Default: \code{energy}}\\
\code{force\_parameter\_name} & \code{dft\_force} & \makecell[tl]{Target force key in the extended XYZ file. \\Default: \code{force}}\\
\code{virial\_parameter\_name} & \code{dft\_virial} & \makecell[tl]{Target virial stress key in the extended XYZ file. \\Default: \code{virial}}\\
\code{e0\_offset} & \code{2.0} & \makecell[tl]{Shifts the baseline energy which is determined\\ from the isolated atom energy.} \\

\code{sparse\_jitter} & \code{1e-8} & Regularisation of $\mathbf{K}_{MM}$ \\

\code{default\_kernel\_regularisation} & \code{\{0.001 0.1 0.05 0.0\}} & \makecell[tl]{Kernel regularisation for target values. Format: \\
\code{\{energy force virial hessian\}}} \\
% \code{config\_type\_kernel\_regularisation} & \{ & energies forces virials hessians \\
% & dimer:0.1:1.0:1.0:0.0: & Override factors for some xyz frames \\
% & hea\_short\_range:0.05:0.8:2.0:0.0: & type energies forces virials hessians \\
% & hea\_surface:0.01:0.4:1.0:0.0: & ... \\
% & isolated\_atom:0.0001:0.04:0.01:0.0: & \\
% & liquid:0.01:0.5:2.0:0.0: & \\
% & liquid\_composition:0.01:0.5:2.0:0.0: & \\
% & liquid\_hea:0.01:0.5:2.0:0.0: & \\
% & short\_range:0.05:0.8:0.8:0.0: & \\
% & surf\_liquid:0.01:0.4:0.2:0.0 & \\
% \} & & \\
\code{config\_type\_kernel\_regularisation} &
\makecell[tl]{ \code{\{}\\
\code{liq:0.003:0.15:0.2:0.0:}\\
\code{amorph:0.01:0.2:0.4:0.0:}\\
\code{sp:0.01:0.2:0.4:0.0:}\\
\code{\}}
} & \makecell[tl]{Override factors for tagged XYZ frames\\format: \\\code{\{config\_type:energies:forces:virials:hessians\}}}\\
\code{gap} & \makecell[tl]{\{
\code{soap}  \\
\code{n\_max=6}  \\
\code{l\_max=6}  \\
\code{atom\_gaussian\_width=0.5}  \\
\code{soap\_exponent=4}  \\
\code{n\_sparse=8000}  \\
\code{cutoff=5.0}  \\
\code{cutoff\_transition\_width=1.0}  \\
\code{sparse\_method=cur\_points}  \\
\code{covariance\_type=dot\_product}  \\
\code{energy\_scale=3.0}  \\
\code{\}}
} & \makecell[tl]{
SOAP descriptor, used as a template.\\
\quad One per element is generated, based\\
\quad on the configurations in the database.\\
Broadening of the atoms in the neighbour density ($\sigma$)\\
Exponent of the polynomial soap kernel \\
Many sparse points are needed due to\\
\quad the high dimensionality of the descriptor.\\
Length scale of radial cutoff in \AA\\
Sparse points are chosen using the CUR method.\\
Form of the kernel\\
Prefactor of the kernel in eV\\
}\\
% & \}} 
%\hline
\end{tabular}
\end{table*}

\added{While keeping all other parameters constant, we individually varied the SOAP parameters $n_{\max}$, $l_{\max}$, $r_{\textrm{cutoff}}$ and $\sigma$, as well as the polynomial kernel exponent $\zeta$ and the number of sparse points $M$.
Based on the original Si \ac{GAP} model, the parameters, $n_{\max}=6$, $l_{\max}=6$, $r_{\textrm{cutoff}}=5 \,\rm{\AA}$, $\sigma=0.5\,\rm{\AA}$, $\zeta=4$ and $M=8000$ were used.}

\added{We have evaluated the interaction energies, forces and virial stresses of all atomic configurations in the test set with the resulting models, and calculated the \ac{RMSE} with respect to the \abinitio{} energies.}

\added{To illustrate how the choice of the parameters affects the computational cost of each model, we have also determined the average calculation time per atom, using a desktop computer utilising a single core of an Intel\textregistered{} Core\texttrademark{} i5-9600K CPU at 3.70~\si{\giga\Hz}.
}

\added{
Trends are presented in Fig.~\ref{fig:Si}, generally showing that more complex models, i.e., those with higher $l_{\max}$, $n_{\max}$, $\zeta$ and $M$ values, are more accurate.
Thanks to the regularisation term in the \ac{GPR}, higher complexity does not result in overfitting.
However, the computational cost of models with higher $l_{\max}$, $n_{\max}$ and sparse points is increased due to the larger number of terms.
Even though the polynomial kernel at higher orders results in more terms, these are not calculated explicitly, therefore the computational cost remains approximately constant at different $\zeta$ values.
}

\added{
Increasing the spatial cutoff of the atomic neighbourhood environment results in more accurate models up to $5\;\textrm{\AA}$, as further neighbours may influence the local energy function.
However, at higher cutoff values the quality of the model deteriorates, which may be regarded as a sign of underfitting, when the available data is not sufficient to determine the dependence of the local energy terms on further neighbours.}

\added{
The cutoff radius of SOAP or other descriptors may also be chosen by considering the force constant matrix of the atomic system, using a criterion on the spatial decay of the elements\cite{Deringer.2020}.
}

\added{
The smoothness parameter $\sigma$ has a strong influence on the accuracy of the model, which is related to how the neighbouring atoms are represented.
Narrow Gaussians lead to fewer similar kernel values between two atomic environments, resulting in overfitting, whereas with wide Gaussians the resolution of the representation is lower.
}

\added{The kernel regularisation hyperparameters provide control on the accuracy and smoothness of the resulting model.
Lower values bias the potential to fit the training data more accurately, but may result in overfitting.
For a given spatial cutoff in the descriptors, the kernel regularisation on the forces may be derived from the decay of the force constant matrix or by quantifying the force uncertainty from \abinitio{} calculations~\cite{Deringer.2020}.
Finding appropriate figures for the kernel regularisation of energy and virial stress values may require cross-validation, but a typical target energy error in condensed systems is 1~\si{\milli\eV/atom}, therefore this is often a suitable starting figure.
The choices of kernel regularisation hyperparameters is discussed extensively in Ref.~\cite{Deringer.2020}.
}

\added{To illustrate the effect of varying the kernel regularisation hyperparameters, we fitted a series of silicon \ac{GAP} models using the same parameters, listed in Table~\ref{tab:configsi}, and same training database as previously, but varied the hyperparameters corresponding to energy ($\sigma_{\textrm{energy}}$), force ($\sigma_{\textrm{force}}$) and virial stress ($\sigma_{\textrm{virial}}$) independently.
We have utilised the test set to predict energy, force and virial stress values using the resulting models, and evaluated \ac{RMSE} figures for each model and quantity.}

\added{
These results are shown in Fig.~\ref{fig:sigma_trend}, highlighting how different choices of regularisation hyperparameters affect the quality of the fit.
For the \ac{RMSE} of the predicted energies, there is an optimal value of $\sigma_{\textrm{energy}}$, while the \ac{RMSE} of the predicted forces and virial stresses decrease monotonically when increasing $\sigma_{\textrm{energy}}$.
Similar opposing changes in the errors of predicted quantities may be observed when $\sigma_{\textrm{force}}$ and $\sigma_{\textrm{virial}}$ are varied.
We attribute these trends to the epistemic uncertainty of our models which is due to our assumptions, such as the locality of the descriptor or the body-order representation.}
\added{These assumptions limit the simultaneous accuracy the model may achieve in energies, forces and virial stresses, and biasing the fit towards reproducing a particular quantity causes a deterioration in others.}

\added{These results are intended to provide a guide to fitting \ac{GAP} models and their adaptation is almost certainly necessary when fitting potentials representing different atomic systems.
While an exhaustive hyperparameter search is not always feasible, the recently parallelised \code{gap\_fit} allows rapid creation and subsequent evaluation of models.}

\subsection{Local Property}\label{sec:localproperty}

\begin{figure}[htb]
\includegraphics[width=\columnwidth]{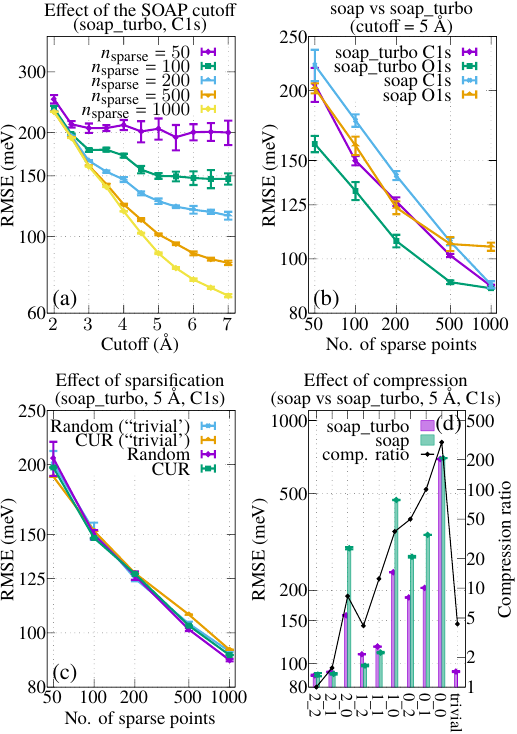} 
\caption{
%Effect of different technical SOAP parameters on the accuracy of different models
%fitted to $GW$ CEBE data for a CHO-containing subset of the QM9 database~\cite{golze_2022}.
(a) RMSEs for a \texttt{soap\_turbo} C1s model as a function of SOAP cutoff and
sparse set size with random selection.
(b) Comparison between \texttt{soap} and \texttt{soap\_turbo} C1s and O1s models
with random sparse set size (fixed cutoff of 5~\AA{}).
(c) Effect of random and CUR sparsification strategies, for a \texttt{soap\_turbo}-based
C1s model (fixed cutoff of 5~\AA{}); the effect of adding compression
(\texttt{soap\_turbo}'s ``trivial'' recipe) is also tested.
(d) Effect of different compression recipes on \texttt{soap} and \texttt{soap\_turbo}
C1s model performance (fixed cutoff of 5~\AA{}); the right axis gives the compression
ratio of the descriptor as the number of dimensions of the full descriptor (2700)
divided by the number of dimensions of the compressed descriptor.
In all cases (a-d), error bars are estimated from the standard deviation of the RMSEs
calculated among 10 different models with random sparse set selection.}
\label{fig:local_property}
\end{figure}

While the local energies predicted by GAPs in cohesive energy models are not physical
observables, there are different local atomic properties with physical significance
amenable to direct learning within the GAP framework. SOAP descriptors are particularly
suited for this task. Examples of such models that we have trained in the past include
adsorption energies~\cite{caro_2018c}, effective Hirshfeld volumes~\cite{muhli_2021b}
and core-electron binding energies (CEBEs)~\cite{golze_2022}.
Here, we revisit the CEBE database of Ref.~\cite{golze_2022} and use it
as a test bed for the performance of SOAP-based local property models as a function of
different convergence parameters.

First, let us briefly provide the context for the usefulness of CEBEs in materials science.
X-ray photoelectron spectroscopy (XPS) uses monochromatic (fixed energy) X-ray light to
excite the deep-lying core electrons in materials and molecules. In oxygen- and
carbon-containing compounds these are the 1s states. When an X-ray photon with sufficient energy
is absorbed by one such core electron the latter becomes photoejected, in such a way that its kinetic
energy can be measured by a detector. Since the energy of the incident photons is fixed,
the difference between the measured kinetic energy and the incident energy equals the CEBE.
This CEBE is characteristic of the chemical environment around the atom whose core electron
was excited, and so an XPS experiment provides a spectrum whose characteristic peaks
give insight into the atomic structure of the material or molecule being probed. Because the
core electron is strongly localized around the nucleus and only feels the influence of the
immediate surrounding medium, XPS is particularly well suited to learning with local
atomic descriptors such as SOAP, as we showed in Ref.~\cite{golze_2022}.
One of the results presented in that paper is a database of $GW$-level CEBEs for a subset of
the CHO-containing molecules in the QM9 dataset~\cite{ramakrishnan_2014}, a large dataset of
small stable organic molecules, containing up to 9 non-hydrogen atoms.

Ref.~\cite{golze_2022} presented learning curves for C1s and O1s CEBE models trained from the
QM9-$GW$ data using \texttt{soap\_turbo} descriptors without sparsification. Here we take a more detailed look at the effect
of different technical parameters on the quality of the fit: cutoff radius for SOAP neighbours,
sparsification scheme, \texttt{soap} vs \texttt{soap\_turbo} and the effect of different
compression recipes on the results. We start out by splitting the QM9-$GW$ CEBE database of
Ref.~\cite{golze_2022} into a training set (80~\% of the structures) and a test set (20~\%
of the structures).

The \texttt{gap\_fit} local property feature relies on the user providing a per-atom local
property array. In this case, an ASE-format XYZ file is provided with a list of per-atom
CEBEs. The following is an example for a formaldehyde molecule:
\begin{widetext}
\begin{verbatim}
4
Lattice="10.89392123 0.0 0.0 0.0 10.81307256 0.0 0.0 0.0 9.01510639" \
Properties=species:S:1:pos:R:3:GW_CEBE:R:1:local_property_mask:L:1 pbc="T T T"
C        5.45500662       5.70744485       4.50565078     294.52870000  T
O        5.47139525       4.50000000       4.50000000     538.69210000  T
H        6.39392123       6.31307256       4.50174003       0.00000000  F
H        4.50000000       6.28730276       4.51510639       0.00000000  F
\end{verbatim}
\end{widetext}
The fifth column contains the 1s CEBE for the C and O atoms, given in eV in this example.
Since H atoms do not have a
core, CEBEs for these atoms are not available, and we pad the array with zeros. A mask is provided to let \texttt{gap\_fit}
know these are to be ignored during the fit.

In addition to the database of atomic structures and observables (CEBEs here), one
needs to provide the name of the local property as specified in the database
(\texttt{local\_property\_parameter\_name="GW\_CEBE"} here), the default regularization
parameter (\texttt{default\_local\_property\_sigma}=0.01 here, in the same units
as the local property), and possibly offsets for the properties to be learned
(\texttt{local\_property0=\{C:290.816456:O:537.946208:H:0\}} here). In our case,
the \texttt{0} offset is computed as the average CEBE of C1s and O1s core electrons
separately. It is important to provide these offsets so that the ML model only needs to
fit the (smooth) differences in the local property values, and not the absolute numbers,
which are significantly harder to learn. Otherwise, the specification of the atomic
descriptors is done in the same way as for a regular GAP model.

All the results for this example are summarized in Fig.~\ref{fig:local_property}. In panel
(a) we show a comparison of models for C1s CEBEs fitted using a \texttt{soap\_turbo}
descriptor with varying cutoff radii and varying number of sparse
points; the cutoff radius is the single
most important hyperparameter in SOAP-based models. Clearly, the accuracy of the models
can be systematically increased by increasing the cutoff. However, the number of sparse
points limits the expressivity of the model, and models with less sparse points will not
benefit from further increasing the cutoff beyond a certain point. E.g., with 50 sparse
points a 3~\AA{} model performs equal to a 7~\AA{} model. We observe that the statistical
variation in the model performance increases with the cutoff, due to the corresponding
increase in the size of configuration space covered by the descriptor (the error bars
are given as the standard deviation computed over 10 different models obtained from 10 different
randomly chosen sparse sets).

In Fig.~\ref{fig:local_property}(b) we show a comparison of C1s and O1s models fitted
using \texttt{soap} and \texttt{soap\_turbo} descriptors of the same dimensionality
(CHO-sensitive descriptors with 8 radial basis functions per element and up to 8th
degree spherical harmonics, the same basis set as used for all the calculations in
Fig.~\ref{fig:local_property}). \texttt{soap\_turbo} models perform slightly better
than \texttt{soap} models, except for the C1s models with maximum number (1000) of
sparse points, where they perform equally.

If Fig.~\ref{fig:local_property}(c) we assess the effect of using random sparse point
selection vs using CUR decomposition to select the sparse set descriptors, as well as the
possible effect of using descriptor compression (\texttt{soap\_turbo}'s ``trivial''
compression recipe) on the results. For this numerical test, the performance of all
models is essentially the same for most practical purposes.

Finally, in Fig.~\ref{fig:local_property}(d) we perform a thorough numerical test of
different compression recipes on the accuracy of the QM9-$GW$ models. The $i\_j$ labeling
convention refers to the $\nu_\text{R} \equiv i$ and $\nu_\text{S} \equiv j$ \texttt{soap}
sensitivity parameters discussed above, and their quasiequivalent recipes for
\texttt{soap\_turbo}. Additionally, \texttt{soap\_turbo} can use the ``trivial'' compression
scheme as detailed above. In addition to the root-mean-squared error (RMSE), the graph shows
the compression ratio computed as the number of dimensions of the compressed descriptor
divided by those of the full descriptor. Unsurprisingly, the errors increase with the compression
ratio, with most compression recipes providing better performance for \texttt{soap\_turbo},
except for 2\_1, 1\_2 and 1\_1, where \texttt{soap} performs slightly better. That is,
the advantage of using \texttt{soap\_turbo} increases with the degree of compression, whereas
\texttt{soap} performs equally or slightly better than \texttt{soap\_turbo} at low compression
ratios. The ``trivial'' compression recipe is only available for \texttt{soap\_turbo} and
provides arguably the best compromise between accuracy and compression ratio (a factor
of 4.3 vs uncompressed SOAP) among the tested
schemes, at least for this particular test with the QM9-$GW$ database.

Although not shown here, the relative advantage of \texttt{soap\_turbo} vs
\texttt{soap} increases when looking at the O1s models, likely because the
QM9-$GW$ training database we used contains significantly more C1s data
(11.5k entries) than O1s data (1.5k entries). This indicates better
generalization and data efficiency for \texttt{soap\_turbo}, although it is
important to note that the performance of each descriptor is dataset
specific.

\subsection{High-Entropy Alloy}

\begin{table*}[!t]
\centering
\caption{Contents of file \code{config} used to fit a \ac{GAP} model for the Mo-Nb-Ta-V-W quinary high-entropy alloy.} 
\label{tab:config}
\begin{tabular}{lll}
\hline
\textbf{Key} & \textbf{Value} & \textbf{Comments} \\
\hline
\code{atoms\_filename} & \code{db\_HEA\_reduced.xyz} & Extended XYZ file of training configurations \\
\code{do\_copy\_at\_file} & \code{F} &  Do not copy XYZ data to output XML \\
\code{gap\_file} & \code{gp\_HEA.xml} & Filename of output XML of \ac{GAP} model\\
\code{sparse\_jitter} & \code{1e-8} & Regularisation of $\mathbf{K}_{MM}$ \\
\code{default\_kernel\_regularisation} & \code{\{0.002 0.1 0.5 0.0\}} & \makecell[tl]{Kernel regularisation for target values\\ format: \code{\{energy force virial hessian\}}} \\
% \code{config\_type\_kernel\_regularisation} & \{ & energies forces virials hessians \\
% & dimer:0.1:1.0:1.0:0.0: & Override factors for some xyz frames \\
% & hea\_short\_range:0.05:0.8:2.0:0.0: & type energies forces virials hessians \\
% & hea\_surface:0.01:0.4:1.0:0.0: & ... \\
% & isolated\_atom:0.0001:0.04:0.01:0.0: & \\
% & liquid:0.01:0.5:2.0:0.0: & \\
% & liquid\_composition:0.01:0.5:2.0:0.0: & \\
% & liquid\_hea:0.01:0.5:2.0:0.0: & \\
% & short\_range:0.05:0.8:0.8:0.0: & \\
% & surf\_liquid:0.01:0.4:0.2:0.0 & \\
% \} & & \\
\code{config\_type\_kernel\_regularisation} &
\makecell[tl]{ \code{\{}\\
\code{dimer:0.1:1.0:1.0:0.0:}\\
\code{hea\_short\_range:0.05:0.8:2.0:0.0:}\\
\code{hea\_surface:0.01:0.4:1.0:0.0:}\\
\code{isolated\_atom:0.0001:0.04:0.01:0.0:}\\
\code{liquid\_composition:0.01:0.5:2.0:0.0:}\\
\code{liquid\_hea:0.01:0.5:2.0:0.0:}\\
\code{short\_range:0.05:0.8:0.8:0.0:}\\
\code{surf\_liquid:0.01:0.4:0.2:0.0}\\
\code{\}}
} & \makecell[tl]{Override factors for tagged XYZ frames\\format: \\\code{\{config\_type:energies:forces:virials:hessians\}}}\\
\code{gap} & \makecell[tl]{\code{\{}\\
\code{distance\_2b} \\
\code{n\_sparse=20} \\
\code{sparse\_method=uniform} \\
\code{covariance\_type=ard\_se} \\
\code{cutoff=5.0} \\
\code{cutoff\_transition\_width=1.0} \\
\code{energy\_scale=10.0} \\
\code{lengthscale\_uniform=1.0} \\
\code{:}} & \makecell[tl]{
Two-body descriptor, used as a template.\\
One per element pair is generated.\\
In one dimension few sparse points\\
(uniformly spaced) are enough.\\
Kernel is squared exponential (se).\\
\\
\\
\\
}\\
& \makecell[tl]{
\code{soap}  \\
\code{n\_sparse=4000}  \\
\code{sparse\_method=cur\_points}  \\
\code{covariance\_type=dot\_product}  \\
\code{n\_max=8}  \\
\code{l\_max=4}  \\
\code{soap\_exponent=2.0}  \\
\code{atom\_gaussian\_width=0.5}  \\
\code{cutoff=5.0}  \\
\code{cutoff\_transition\_width=1.0}  \\
\code{energy\_scale=1.0}  \\
\code{\}}
} & \makecell[tl]{
SOAP descriptor, used as a template.\\
One per element is generated, based\\
on the configurations in the database.\\
\\
Many sparse points are needed due to\\
the high dimensionality of the descriptor.\\
\\
Sparse points are chosen using the CUR method.\\
}\\
% & \}} 
%\hline
\end{tabular}
\end{table*}

The Mo-Nb-Ta-V-W quinary high-entropy alloy studied by Byggm\"astar \etal{}\cite{Byggmastar.2021} is a complex, multicomponent system.
To illustrate the generation of a \ac{GAP} model, we provide the parameters, complete with explanations and comments, that were used to test and benchmark the \ac{MPI}-{ScaLAPACK} implementation of the \code{gap\_fit} program\cite{Klawohn.2023}.
\added{Here we highlight a recent feature which conveniently allows the fitting parameters to be stored in a file, rather than provided as command line arguments.}
%This training set was used for the performance evaluation of the MPI/ScaLAPACK implementation.

The fitting parameters are entered in the file \code{config}, provided in the supplementary information\cite{si}, in a \code{key=value} format, with commentary on each provided in Table~\ref{tab:config}.
The fitting database is openly available as the supplementary material of the article by Byggm\"astar \etal{}\cite{Byggmastar.2021}, which may be downloaded from the Fairdata repository\cite{HEA}.

The fitting procedure can then be carried out by running the command \code{gap\_fit config\_file=config} with the database file \code{db\_HEA\_reduced.xyz} and the configuration file \code{config} in the same directory.
For the parallel implementation, the command line \code{mpirun -np 2 gap\_fit config\_file=config} executes the process on two computational cores.
Hybrid OpenMP-\ac{MPI} execution is possible, for which the number of threads may be adjusted by setting the environment variable as \code{export OMP\_NUM\_THREADS=4}.
For the specific queuing system available to the user, the documentation should be consulted.

The resulting \ac{GAP} model is stored in the \code{gp\_HEA.xml} file and a set of text files according to the naming pattern \code{gp\_HEA.xml.sparseX.GAP\_*}.
The interatomic potential may be accessed as a \code{Calculator} in ASE as
\begin{verbatim}
from quippy.potential import Potential
p = Potential(param_filename="gp_HEA.xml")
...
a.calc = p
\end{verbatim}
\added{where the Python variable \code{a} indicates an ASE \code{Atoms} object.}

\added{Massively parallel simulations with \ac{GAP} models are possible with LAMMPS.
To use the high-entropy alloy potential, the following lines should be added to the LAMMPS input file:}
\begin{verbatim}
pair_style  quip
pair_coeff  * * gp_HEA.xml "" 42 41 73 23 74
\end{verbatim}
\added{where the LAMMPS atom types 1, 2, 3, 4, and 5 are mapped to Mo, Nb, Ta, V and W, respectively.}
\deleted{a \code{pair\_style} in LAMMPS.}

% Download the xyz file from \cite{Byggmastar:2021}, create a file \code{config} in the same directory and paste the left column of the following page into it. Some annotations are given in the right column. The fit can then be run with \code{gap\_fit config\_file=config}.

% If you have compiled with OpenMP, you can set the number of threads used with \code{export OMP\_NUM\_THREADS=4}.

% If you have compiled with MPI/ScaLAPACK you can use \code{mpirun -np 2 gap\_fit config\_file=config} to run with two processes. Using a queuing system like Slurm, running on multiple nodes can be as easy as \code{srun gap\_fit config\_file=config} where the number of nodes, tasks per node and CPUs per task are set either with additional \code{srun} options or \code{\#SBATCH} directives in a job script.

%\textcolor{red}{(Couldn't get the annotated config example to work within the normal text flow. Switching to onecolumn does not seem to work. Neither does using a float.)}
% J. Byggm ̈astar, K. Nordlund, and F. Djurabekova, Machine-learned interatomic potential for Mo-Nb-Ta-V-W (2+3-body tabGAP) (2021)
% \url{https://doi.org/10.23729/1b845398-5291-4447-b417-1345acdd2eae}

\section{Conclusion and outlook}

We have reviewed the \ac{GAP} framework from an implementation point of view, highlighting how a generic sparse \ac{GPR} formalism is adapted for the prediction of interatomic potentials and related quantities.
An overview of the software package, coding practices and recent developments was provided, together with usage examples and detailed explanations of adjustable parameters, with references to the theory.
The \ac{QUIP} and \ac{GAP} suite remains under maintenance and in active development by the authors, and will serve as a test bed for exploring ideas in the field of \acp{MLIP}.
With interfaces to Python and major simulation packages, \ac{GAP} serves as a useful tool for computational modelling.

Future developments will include the inclusion of more descriptors, such as \ac{ACE}\cite{ACE_ralf}, rigorous means for uncertainty quantification, utilising modern computing architectures such as GPUs, and the implementation of more robust and efficient solvers for the fitting procedure.
With the availability and popularity of more modern programming languages such as Python or Julia, Fortran may appear as an outdated choice.
However, its interoperability with \ac{MPI} and related libraries such as ScaLAPACK has proved to be an advantage in utilising the more traditional, but still highly prevalent, high performance computing facilities consisting of networked servers.
As supercomputers and programming skills change, the current framework might prove to be too restrictive, but the practical insight documented here and the software will remain valuable for future endeavours.

\section*{Acknowledgments}
This work was financially supported by the NOMAD Centre of Excellence (European Commission grant agreement ID 951786) and the Leverhulme Trust Research Project Grant (RPG-2017-191).
ABP acknowledges support from the CASTEP-USER project, funded by the Engineering and Physical Sciences Research Council under the grant agreement EP/W030438/1. MAC acknowledges personal
funding from the Academy of Finland under grant \#330488.
We acknowledge computational resources provided by the Max Planck Computing and Data Facility provided through the NOMAD CoE, the Scientific Computing Research Technology Platform of the University of Warwick, the EPSRC-funded HPC Midlands+ consortium (EP/T022108/1), ARCHER2 (https://www.archer2.ac.uk/) via the UK Car-Parinello consortium (EP/P022065/1), CSC - IT Center for Science, and the Aalto University Science-IT project. We thank the technical staff at each of these HPC centres for their support.
%shown the concept of Gaussian Approximation Potentials (GAPs) and their implementation in Gap Fit within our QUIP software package.

\section*{Author declarations}
\subsection*{Conflict of interest}
ABP and GC are listed as inventors on a patent filed by Cambridge Enterprise Ltd. related to SOAP and GAP (US patent 8843509, filed on 5 June 2009 and published on 23 September 2014). ABP, MAC and GC benefit from licensing the \ac{GAP} software to industrial users. Not-for-profit use for academic and educational purposes is granted under the Academic Software License for no cost. The other authors have no conflicts to disclose.

\subsection*{Author Contributions}
\textbf{Sascha Klawohn}: data curation (lead); methodology (supporting); formal analysis (equal); validation (supporting); visualization (equal); writing (supporting).
\textbf{James P. Darby}: data curation (lead); methodology (supporting); formal analysis (supporting); validation (supporting); visualization (equal); writing (supporting).
\textbf{James R. Kermode}: methodology (supporting); supervision (equal); funding acquisition
(equal); writing (supporting).
\textbf{G\'abor Cs\'anyi}: conceptualization (equal); methodology (equal); supervision (equal); funding acquisition
(equal); writing (supporting).
\textbf{Miguel A. Caro}: conceptualization (support); formal analysis (supporting); validation (supporting); visualization (equal); methodology (equal); writing (supporting).
\textbf{Albert P. Bart\'ok}: conceptualization (equal); formal analysis (equal); validation (equal); methodology (equal); supervision (equal); funding acquisition (equal); writing (lead).

\subsection*{Data availability}
All databases referred to this work are available in public data repositories.

\bibliographystyle{apsrev4-2}
\bibliography{gap_fit}% Produces the bibliography via BibTeX.

\appendix

\section*{Arguments of \code{gap\_fit}}
\label{sec:args}

%\todo{Add options for selected descriptors, currently distance 2b and soap}
%\todo[inline]{Group options, e.g. data-targeted (atoms filename, parameter names), or use alternative groupings with just a list of the options, their help strings can then be looked up in the alphabetically ordered section.}

We provide a snapshot of the currently available command line arguments of the \code{gap\_fit} program for completeness.
As the \ac{QUIP} and \code{GAP} packages are under constant development, this list may change.
It should also be noted that most of the parameters need little adjustment, whereas those that pertain to the particular fitting problem are mandatory, requiring the user to specify a value.
Some keywords have aliases, often less descriptive, but both of which are acceptable. These are indicated as \code{option \textcolor{gray}{(alias)}}.

\subsection{Common arguments}

\begin{arglist}
    \setlength\itemsep{0em}
    \item[config\_file] File as alternative input to command line arguments. Newlines are converted to spaces.
    \item[atoms\_filename \textcolor{gray}{(at\_file)}] extended XYZ file containing database configurations in a concatenated form
    \item[gap] Initialisation string for GAPs
    \item[e0] Atomic energy value to be subtracted from energies before fitting, and added back on after prediction. Possible options are: a single number, used for all species; or by species, e.g.: \code{\{Ti:-150.0:O:-320.1\}}.
    \item[local\_property0] Local property value to be subtracted from the local property before fitting, and added back on after prediction. Possible options are: a single number, used for all species; or by species: e.g. \code{\{H:20.0:Cl:35.0}\}.
    \item[e0\_offset] Offset of baseline. If zero, the offset is the average atomic energy of the input data or the e0 specified manually.
    \item[e0\_method] Method to determine the constant energy baseline \code{e0}, if not explicitly specified. Possible options: \code{isolated} (default, each atom present in the XYZ needs to have an isolated representative, with a valid energy); \code{average} (e0 is the average of all total energies across the XYZ).
    \item[default\_kernel\_regularisation \textcolor{gray}{(default\_sigma)}] Prior assumption of error in [energies forces virials hessians]
    \item[\parbox{\linewidth}{default\_kernel\_regularisation\_local\_property \\
    \textcolor{gray}{(default\_local\_property\_sigma)}}] Prior assumption of error in \code{local\_property}.
    \item[sparse\_jitter] Extra regulariser used to regularise the sparse covariance matrix before it is passed to the linear solver. Use something small, it really shouldn't affect your results, if it does, your sparse basis is still very ill-conditioned.
    \item[hessian\_displacement \textcolor{gray}{(hessian\_delta)}] Finite displacement to use in numerical differentiation when obtaining second derivative for the Hessian covariance.
    \item[baseline\_param\_filename \textcolor{gray}{(core\_param\_file)}] QUIP XML file which contains a potential to subtract from data, and added back after prediction.
    \item[baseline\_ip\_args \textcolor{gray}{(core\_ip\_args)}]  QUIP initialisation string for a potential to subtract from data, and added back after prediction.
    \item[energy\_parameter\_name] Name of energy property in the input extended XYZ file that describes the data
    \item[local\_property\_parameter\_name] Name of \code{local\_property} as a column in the input XYZ file that describes the data.
    \item[local\_property\_mask\_parameter\_name] Used to exclude local properties on specific atoms from the fit. In the XYZ, it must be a logical column.
    \item[force\_parameter\_name] Name of force property, as three columns, in the input XYZ file that describes the data.
    \item[virial\_parameter\_name] Name of virial property in the input XYZ file that describes the data.
    \item[stress\_parameter\_name] Name of stress property (6-vector or 9-vector) in the input XYZ file that describes the data - stress values only used if virials are not available. Note the opposite sign and standard Voigt order.
    \item[hessian\_parameter\_name] Name of hessian property (column) in the input XYZ file that describes the data
    \item[config\_type\_parameter\_name] Allows grouping on configurations into. This option is the name of the key that indicates the configuration type in the input XYZ file. With the default, the key-value pair \code{config\_type=bcc} would place that configuration into the group \code{bcc}.
    \item[\parbox{\linewidth}{kernel\_regularisation\_parameter\_name \\ 
    \textcolor{gray}{(sigma\_parameter\_name)}}] Kernel regularisation parameters for a given configuration in the database. Overrides the command line values for both defaults and config-type-specific values. In the input XYZ file, it must be prepended by \code{energy\_}, \code{force\_}, \code{virial\_} or \code{hessian\_} keywords.
    \item[force\_mask\_parameter\_name] To exclude forces on specific atoms from the fit. In the XYZ, it must be a logical column.
    \item[parameter\_name\_prefix] Prefix that gets uniformly appended in front of \code{\{energy,local\_property,force,virial,...\}\\ \_parameter\_name}
    \item[\parbox{\linewidth}{config\_type\_kernel\_regularisation \\ 
    \textcolor{gray}{(config\_type\_sigma)}}] The kernel regularisation values to choose for each type of data, when the configurations are grouped into config\_types. Format: \code{\{configtype1:energy:force:virial:hessian:\\ config\_type2:energy:force:virial:hessian\}}
    \item[\parbox{\linewidth}{kernel\_regularisation\_is\_per\_atom \\ 
    \textcolor{gray}{(sigma\_per\_atom)}}] Interpretation of the energy and virial regularisation parameters specified in \code{default\_kernel\_regularisation} and \code{config\_type\_kernel\_regularisation}. If \code{T}, these are interpreted as per-atom errors, and the variance will be scaled according to the number of atoms in the configuration. If \code{F}, they are treated as absolute errors and no scaling is performed. NOTE: values specified on a per-configuration basis (see \code{kernel\_regularisation\_parameter\_name}) are always absolute, not per-atom.
    \item[do\_copy\_atoms\_file \textcolor{gray}{(do\_copy\_at\_file)}] Copy the input XYZ file into the GAP XML file (should be set to False for NetCDF input).
    \item[sparse\_separate\_file] Save sparse point data in separate file, recommended for large number of sparse points.
    \item[sparse\_use\_actual\_gpcov] Use actual GP covariance for sparsification methods.
    \item[gap\_file \textcolor{gray}{(gp\_file)}] Name of output XML file that will contain the fitted potential
    \item[verbosity] Verbosity control. Options: \code{NORMAL}, \code{VERBOSE}, \code{NERD}, \code{ANALYSIS}.
    \item[rnd\_seed] Random seed.
    \item[openmp\_chunk\_size] Chunk size in OpenMP scheduling.
    \item[do\_ip\_timing] To enable or not the timing of the interatomic potential.
    \item[template\_file] Template XYZ file for initialising object.
    \item[sparsify\_only\_no\_fit] If true, sparsification is done, but no fitting is performed. The sparse index is printed by adding \code{print\_sparse\_index=file.dat} to the descriptor specification string under the \code{gap} option.
    \item[dryrun] If true, exits after memory estimate, before major allocations.
    \item[condition\_number\_norm] Norm for condition number of matrix A of the linear system; \code{O}: 1-norm, \code{I}: inf-norm, \code{<empty>}: skip calculation (default)
    \item[linear\_system\_dump\_file] Basename prefix of linear system dump files. Skipped if \code{<empty>} (default).
    \item[mpi\_blocksize\_rows] Blocksize of MPI distributed matrix rows. Affects efficiency and memory usage slightly. Maximum if specified as 0 (default).
    \item[mpi\_blocksize\_cols] Blocksize of MPI distributed matrix columns. Affects efficiency and memory usage considerably. Maximum if 0. Default: 100.
    \item[mpi\_print\_all] If true, each MPI processes will print its output. Otherwise, only the first process does (default).
\end{arglist}

\subsection*{Arguments of the GAP string}
The following keywords are to be specified for each descriptor within the \code{gap} command line argument.
\begin{arglist}
    \setlength\itemsep{0em}
    \item[energy\_scale \textcolor{gray}{(delta)}] Set the typical scale of the function being fitted, or the specific energy term if using multiple descriptors. It is equivalent to the standard deviation of the Gaussian Process in the probabilistic view, and typically this would be set to the standard deviation (i.e. root mean square) of the function that is approximated with the Gaussian Process.
    \item[f0] Set the mean of the Gaussian Process. Defaults to 0.
    \item[n\_sparse] Number of sparse points to use in the sparsification of the Gaussian Process
    \item[config\_type\_n\_sparse] Number of sparse points in each configuration type. Format: \code{type1:50:type2:100}
    \item[sparse\_method] Sparsification method. Possible options: \code{RANDOM}(default), \code{PIVOT}, \code{CLUSTER}, \code{UNIFORM}, \code{KMEANS}, \code{COVARIANCE}, \code{NONE}, \code{FUZZY}, \code{FILE}, \code{INDEX\_FILE}, \code{CUR\_COVARIANCE}, \code{CUR\_POINTS}. For explanations, see below.
    \item[lengthscale\_factor \textcolor{gray}{(theta\_fac)}] Set the width of Gaussians for the Gaussian and piecewise polynomial kernel by multiplying the range of each descriptor by \code{lengthscale\_factor}. Can be a single number or different for each dimension. For multiple numbers in \code{lengthscale\_factor}, separate each value by whitespaces.
    \item[lengthscale\_uniform \textcolor{gray}{(theta\_uniform)}] Set the width of Gaussians for the Gaussian and piecewise polynomial kernel, same in each dimension.
    \item[lengthscale\_file \textcolor{gray}{(theta\_file)}] Set the width of Gaussians for the Gaussian kernel from a file. There should be as many real numbers as the number of dimensions, in a single line.
    \item[sparse\_file] Sparse points from a file. If \code{sparse\_method=FILE}, descriptor values as real numbers listed in a text file, one element per line. If \code{sparse\_method=INDEX\_FILE}, 1-based index of sparse points, one per line.
    \item[mark\_sparse\_atoms] If true, reprints the original extended XYZ file after sparsification process, with a \code{sparse\_property} column added, which is true for atoms associated with a sparse point.
    \item[add\_species] If true (default), create species-specific descriptors, using the descriptor string as a template.
    \item[covariance\_type] Type of covariance function to use. Available: \code{GAUSSIAN}, \code{DOT\_PRODUCT}, \code{BOND\_REAL\_SPACE}, \code{PP} (piecewise polynomial).
    \item[soap\_exponent \textcolor{gray}{(zeta)}] Exponent of \code{soap} type dot product covariance kernel
    \item[print\_sparse\_index] If given, after determining the sparse points, their 1-based indices are appended to this file.
    \item[unique\_hash\_tolerance] Hash tolerance when filtering out duplicate data points.
    \item[unique\_descriptor\_tolerance] Descriptor tolerance when filtering out duplicate data points.
\end{arglist}

\subsection*{Options for sparse point selection}

\begin{arglist}
    \setlength\itemsep{0em}
    \item [none] No sparsification, selects all datapoints.
    \item [index\_file] Reads indices of sparse points from the file given by \code{sparse\_file} and selects those from the de-duplicated data.
    \item [file] Reads sparse points from the file given by \code{sparse\_file}.
    \item [random] Selects \code{n\_sparse} random descriptors with the same probability.
    \item [uniform] Computes a histogram of the data with \code{n\_sparse} bins and returns a data point from each bin. This option is only suitable for low-dimensional descriptors.
    
    \item [kmeans] The $k$-means clustering algorithm is performed on all descriptors to generate \code{n\_sparse} clusters, of which the descriptors closest to the cluster means are selected as sparse points. 
    \item [fuzzy] A fuzzy version of $k$-means clustering\cite{Doring.2006} is used to generate \code{n\_sparse} clusters.
    \item [cluster] A $k$-medoid clustering based on the full covariance matrix of descriptors is performed, resulting in \code{n\_sparse} clusters. The medoid points are selected as sparse points.
    \item [pivot] The \code{n\_sparse} ``pivot'' indices of the full covariance matrix are found, and used as the sparse points.
    \item [covariance] Greedy data point selection based on the sparse covariance matrix, to minimise the \ac{GPR} variance of all datapoints.
    \item [cur\_points] A CUR decomposition, based on the datapoints, is carried out to find the most representative \code{n\_sparse} points.
    \item [cur\_covariance] A CUR decomposition, based on the full covariance matrix, is carried out to find the most representative \code{n\_sparse} points.
\end{arglist}

\subsection*{Descriptors}

The \ac{GAP} module implements over 30 descriptors, most of which being experimental or unsupported.
In the following, we include those that are commonly used by practitioners and supported by the \ac{GAP} developers.

% \begin{arglist}
%     \item[A2\_dimer]
%     \item[AB\_dimer]
%     \item[alex]
%     \item[AN\_monomer]
%     \item[angle\_3b]
%     \item[as\_distance\_2b]
%     \item[atom\_real\_space]
%     \item[behler]
%     \item[bispectrum\_so3]
%     \item[bispectrum\_so4]
%     \item[bond\_real\_space]
%     \item[co\_angle\_3b]
%     \item[co\_distance\_2b]
%     \item[com\_dimer]
%     \item[coordination]
%     \item[cosnx]
%     \item[distance\_2b]
%     \item[distance\_Nb]
%     \item[general\_dimer]
%     \item[general\_monomer]
%     \item[general\_trimer]
%     \item[molecule\_lo\_d]
%     \item[power\_so3]
%     \item[power\_so4]
%     \item[rdf]
%     \item[soap\_express]
%     \item[soap\_turbo]
%     \item[soap]
%     \item[trihis]
%     \item[water\_dimer]
%     \item[water\_monomer]
% \end{arglist}

\subsubsection*{\code{distance\_2b} arguments}

\begin{arglist}
    \item[cutoff] Cutoff for \code{distance\_2b}-type descriptors.
    \item[cutoff\_transition\_width] Transition width of cutoff for \code{distance\_2b}-type descriptors.
    \item[Z1] Atom type \#1 in bond. Any atom type if missing.
    \item[Z2] Atom type \#2 in bond. Any atom type if missing.
    \item[resid\_name] Name of an integer property in the atoms object giving the residue identifier of the molecule to which the atom belongs.
    \item[only\_intra] Only calculate bonds, i.e. \emph{intra}molecular pairs with equal residue identifiers
    \item[only\_inter] Only apply to non-bonded atom pairs, i.e. \emph{inter}molecular pairs with different residue identifiers.
    \item[n\_exponents] Number of exponents.
    \item[exponents] Exponents in a list format, for example: \code{\{-12 -6\}}
    \item[tail\_range] Tail range.
    \item[tail\_exponent] Tail exponent.
\end{arglist}

\subsubsection*{\code{soap} arguments}

\begin{arglist}
    \item[cutoff] Cutoff distance.
    \item[cutoff\_transition\_width] Transition width of cutoff function.
    \item[cutoff\_dexp] Cutoff decay exponent.
    \item[cutoff\_scale] Cutoff decay scale.
    \item[cutoff\_rate] Inverse cutoff decay rate.
    \item[l\_max] $l_{\max}$ (spherical harmonics basis band limit) for soap-type descriptors.
    \item[n\_max] $n_{\max}$ (number of radial basis functions) for soap-type descriptors.
    \item[atom\_gaussian\_width \textcolor{gray}{(atom\_sigma)}] Width of atomic Gaussian functions for soap-type descriptors.
    \item[central\_weight] Weight of central atom in environment.
    \item[central\_reference\_all\_species] Place a Gaussian reference for all atom species densities. By default (\code{F}) only consider when neighbour is the same species as centre.
    \item[average] Whether to calculate averaged SOAP - one descriptor per atoms object. If false (default), atomic SOAP is returned.
    \item[diagonal\_radial] Only return the $n_1=n_2$ elements of the power spectrum.
    \item[covariance\_sigma0] $\sigma_0$ parameter in polynomial covariance function.
    \item[normalise \textcolor{gray}{(normalize)}] Normalise descriptor, so magnitude is 1. In this case the kernel of two equivalent environments is 1.
    \item[basis\_error\_exponent]  $10^{-\texttt{basis\_error\_exponent}}$ is the max difference between the target and the expanded function.
    \item[n\_Z] How many different types of central atoms to consider.
    \item[n\_species] Number of species for the descriptor.
    \item[species\_Z] Atomic number of species.
    \item[xml\_version] Version of GAP the XML potential file was created.
    \item[Z] Atomic number of central atom, 0 is the wild-card \emph{or} Atomic numbers to be considered for central atom, must be a list.
    % \item[radial\_basis] Radial basis functions to use. Options are EQUISPACED_GAUSS, POLY and GTO (default for xml_version > 1987654320
\end{arglist}

\subsubsection*{\code{soap} compression arguments}
\begin{arglist}
    \item[Z\_mix] Mix the element channels together if present.% \code{T/F}
    \item[R\_mix] Mix the radial channels together if present. % \code{T/F}
    \item[sym\_mix] Specifies whether a single set of mixing weights is used \emph{or} whether two sets are used. If \code{sym\_mix=T}, tensor-decomposition is enabled. If \code{sym\_mix=F} tensor-sketching is used. 
    \item[K] Integer specifying how many mixed channels to create. For example, \code{R\_mix=T Z\_mix=T K=5} will create 5 mixed channels whereas \code{R\_mix=F n\_max=6 Z\_mix=T K=5} will result in \code{K*n\_max} = 30 channels.
    \item[coupling] If \code{coupling=T} full tensor-product coupling is applied across the resulting channels after mixing, whereas if \code{coupling=F} element-wise coupling is applied instead. The \textbf{only} exception to this rule occurs for \code{Z\_mix=T R\_mix=F} (or similarly for \code{Z\_mix=F R\_mix=T})  with \code{coupling=F}. Here, element-wise coupling is applied across the mixed-element channels but tensor-product coupling is applied across the unmixed radial channels, resulting in $p^k_{nn'l}$ (or similarly $p^{\alpha\beta}_{kl}$). 
    \item[mix\_shift] Integer specifying the shift to the default seed that is used for the random number generator used to generate the mixing weights. 
     \item[nu\_R] radially sensitive correlation order. Allowed values are 0, 1 and 2 (default).
    \item[nu\_S] species sensitive correlation order.  Allowed values are 0, 1 and 2 (default).
    \item[Z\_map]  Commas separate groups within a density. A colon separates the two densities if present. Otherwise the groups are taken to be equal. \texttt{Z\_map}= \code{\{1, 3, 22 23 24 \}} has a separate channel for H and Li but treats Ti, V and Cr as identical
    \texttt{Z\_map} = \code{\{1, 3, 22, 23, 24 : 1, 3, 22 23 24\}} has a separate channel for each element in the first density. In the second density there is a separate channel for H and Li but Ti, V and Cr are identical
\end{arglist}

\subsubsection*{\code{soap\_turbo} arguments}

\begin{arglist}
    \item[rcut\_hard] Hard cutoff distance.
    \item[rcut\_soft] Soft cutoff distance.
    \item[n\_species] Number of species for the descriptor.
    \item[l\_max] $l_{\max}$ (spherical harmonics basis band limit) for soap-type descriptors.
    \item[nf] Sets the rate of decay of the atomic density in the region between soft and hard
    cutoffs.
    \item[radial\_enhancement] Integer index (0, 1 or 2) that simulates the effect of modulating
    the radial overlap integral with the radial distance raised to this number.
    \item[basis] Options: \code{poly3} or \code{poly3gauss} chooses a 3rd and higher degree
    polynomial radial basis set and augments it with a Gaussian at the origin, respectively.
%    \item[scaling\_mode] 
    \item[compress\_file] Optional user-provided file specifying the compression recipe.
    \item[compress\_mode] Optionally provides a predefined compression recipe.
    \item[central\_index] 1-based index of central atom \code{species\_Z} in the species array.
    \item[alpha\_max] Radial basis resolution for each species.
    \item[atom\_sigma\_r] Width of atomic Gaussian functions for soap-type descriptors in the radial direction.
    \item[atom\_sigma\_r\_scaling] Scaling rate of radial sigma: scaled as a function of neighbour distance.
    \item[atom\_sigma\_t] Width of atomic Gaussian functions for soap-type descriptors in the angular direction.
    \item[atom\_sigma\_t\_scaling] Scaling rate of angular sigma: scaled as a function of neighbour distance.
    \item[amplitude\_scaling] Scaling rate of amplitude: scaled as an inverse function of neighbour distance.
    \item[central\_weight] Weight of central atom in environment.
    \item[species\_Z] Atomic number of species, including the central atom.
\end{arglist}

\end{document}